\documentclass{aa}
\usepackage{graphicx}
\usepackage{txfonts}
\usepackage{natbib}
\usepackage[figuresright]{rotating}
\bibpunct{(}{)}{;}{a}{}{,}
\newcommand{\Cr}[5]{\mbox{$#1\,^#2{\rm #3}^{{\rm #4}}_{\rm #5}$}}
\newcommand{\Teff}{T_{\rm eff}}
\newcommand{\Te}{T_{\rm e}}
\newcommand{\EW}{W_{\lambda}}
\newcommand{\mA}{{\rm m\AA}}
\newcommand{\Elow}{E_{\rm low}}
\newcommand{\Vmic}{\xi_{\rm t}}
\newcommand{\Vmac}{V_{\rm mac}}
\newcommand{\kms}{km s$^{-1}$}
\newcommand{\SH}{S\!_{\rm H}}           
\newcommand{\opd}{\log \tau_{\rm c}}
\newcommand{\loggfe}{\log (gf\varepsilon)}
\newcommand{\loggfestar}{\log (gf\varepsilon)^{\ast}}
\newcommand{\loggfesun}{\log (gf\varepsilon)^{\odot}}
\newcommand{\loge}{\log\varepsilon}
\begin{document}
\title{Chromium: NLTE abundances in metal-poor stars and nucleosynthesis in the
Galaxy}
\author{M. Bergemann$^{1}$\thanks{E-mail: mbergema@mpa-garching.mpg.de} and
G. Cescutti$^{2}$\thanks{E-mail: cescutti@oats.inaf.it}} 
\institute{$^{1}$Max-Planck Institute for Astrophysics, Karl-Schwarzschild Str.
1, 85741, Garching, Germany \\
$^{2}$Astronomy Unit, Department of Physics, University of Trieste, via G.B.
Tiepolo 11, I-34143, Trieste, Italy}
\date{Received date / Accepted date}

\abstract
{}
{We investigate statistical equilibrium of Cr in the atmospheres of late-type
stars to show whether the systematic abundance discrepancy between Cr I and Cr
II lines, as often encountered in the literature, is due to deviations from LTE.
Furthermore, we attempt to interpret the NLTE trend of [Cr/Fe] with [Fe/H] using
chemical evolution models for the solar neighborhood.}
{NLTE calculations are performed for the model of Cr atom, comprising $340$
levels and $6806$ transitions in total. We make use of the quantum-mechanical
photoionization cross-sections of Nahar (2009) and investigate sensitivity of
the model to uncertain cross-sections for H I collisions. NLTE line formation is
performed for the MAFAGS-ODF model atmospheres of the Sun and 10 metal-poor
stars with $-3.2 <$ [Fe/H] $< -0.5$, and abundances of Cr are derived by
comparison of the synthetic and observed flux spectra.}
{We achieve good ionization equilibrium of Cr for the models with different
stellar parameters, if inelastic collisions with H I atoms are neglected. The
solar NLTE abundance based on Cr I lines is $5.74$ dex with $\sigma = 0.05$ dex;
it is $\sim 0.1$ higher than the LTE abundance. For the metal-poor stars, the
NLTE abundance corrections to Cr I lines range from $+0.3$ to $+0.5$ dex.
The resulting [Cr/Fe] ratio is roughly solar for the range of metallicities
analyzed here, which is consistent with current views on production of these
iron peak elements in supernovae.}
{The tendency of Cr to become deficient with respect to Fe in metal-poor
stars is an artifact due to neglect of NLTE effects in the line formation of
\ion{Cr}{i}, and it has no relation to peculiar physical conditions in the
Galactic ISM or deficiencies of nucleosynthesis theory. }
\titlerunning{Cr in metal-poor stars}
\authorrunning{M. Bergemann \& G. Cescutti}
\maketitle
\section{Introduction}
Abundance of chemical elements in the atmospheres of late-type stars is a key
information in studies of Galactic chemical evolution (GCE).
Combinations of different elements and variation of their abundances with
metallicity are commonly used to calibrate GCE models and to constrain
poorly-known parameters, like star formation history, initial mass function,
and efficiency of mixing in the ISM. Since stellar yields are also a part of
such models, the abundances can also probe the theories of stellar
nucleosynthesis and evolution and highlight problems in them.

The abundances are derived by means of high-resolution spectroscopy that 
implies modelling the observed stellar spectrum. Whereas a poor quality of a
spectrum introduces some random noise about a true value of the abundance, major \emph{systematic} errors result from an oversimplified treatment of radiation
transfer and convection in stellar atmospheres \citep{2005ARA&A..43..481A}. The
assumption of local thermodynamic equilibrium (LTE), which is traditionally used
to compute a spectrum in an attempt to avoid numerical difficulties with line
formation, breaks down for the many species. Minority ions, which
constitute at most few percent of the total element abundance, are particularly
affected by NLTE conditions. As a consequence, various abundance indicators of
the same element, like lines of different ionization stages or excitation
potentials, often give discrepant results in 1D LTE analysis. The main concern
is that once observations compel us to restrict the analysis to a \emph{single
indicator}, e.g. due to a moderate signal-to-noise ratio or limited spectral
range, spurious abundance trends with metallicity are unavoidable.

For Cr, even-$Z$ element of the Fe-group, only LTE calculations have
been performed up to now. They revealed severe problems with modelling
excitation and ionisation balance of Cr in the atmospheres of late-type stars.
Systematic differences of $0.1 - 0.5$ dex between abundances based on LTE
fitting of the \ion{Cr}{i} and \ion{Cr}{ii} lines were reported for metal-poor
giants and dwarfs \citep{2002ApJS..139..219J, 2003A&A...404..187G,
2008ApJ...681.1524L, 2009A&A...501..519B}. The discrepancies are smaller in the
atmospheres with larger metal content, amounting to $\sim0.1$ dex for Galactic
disk stars \citep{2000AJ....120.2513P} and for the Sun
\citep{2007ApJ...667.1267S,2009ARA&A..47..481A}. Following
\citet{1991A&A...241..501G}, these offsets are usually attributed to the
overionization of \ion{Cr}{i}, a typical NLTE phenomenon affecting minority
atoms in stellar atmospheres. Alternatively, \citet{2004A&A...425..671B}
suggested uneven distribution of neutral and ionized Cr in different atmospheric
layers, which depends on stellar parameters.

As a result, two views on the evolution of Cr abundances in the Galaxy exist in
the literature. The constant [Cr/Fe] with metallicity is derived from the LTE
analysis of \ion{Cr}{ii} lines \citep{1991A&A...241..501G, 2009A&A...501..519B}
that has a simple interpretation in the theory of nucleosynthesis. Cr is formed
with Fe in explosive Si-burning that occurs in supernova (SN) events, and the
production ratio Cr/Fe is roughly solar in both SNe II and SNe Ia \citep{2003hic..book.....C}. Thus, using standard prescriptions for Cr
nucleosynthesis, most of the GCE models reproduce the flat [Cr/Fe] trend with
[Fe/H] \citep[e.g.][]{1995ApJS...98..617T,1998ApJ...496..155S,2000A&A...359..191G}.

On the other hand, most LTE analysis of \ion{Cr}{i} lines
\citep{1995AJ....109.2757M,2004A&A...416.1117C,2004ApJ...612.1107C} indicate 
that the [Cr/Fe] ratio steadily declines towards lowest metallicity. There is
no simple explanation to this trend. Some chemical evolution studies tried
to overcome the problem by introducing correction factors to the theoretical SN 
yields \citep{2004A&A...421..613F,2010ApJ...709..715H}. The assumption
of erroneous supernova yields is not unrealistic, given their sensitivity to the
details of explosion \citep{1999ApJS..125..439I,1999ApJ...517..193N}. However,
to reproduce an approximate trend given by spectroscopic data arbitrary scaling
factors to stellar yields are usually chosen, which lack any physical
justification. Also, recent studies of metal-free massive stars and their
nucleosynthesis yields demonstrate that subsolar Cr/Fe abundance ratios in very
metal-poor stars can not be reproduced by any combination of SN II model
parameters \citep{2005ApJ...619..427U, 2008arXiv0803.3161H}, especially when
other Fe-peak elements are taken into account.

In this paper, we report NLTE abundances of Cr for the Sun and a sample of
dwarfs and subgiants with $-3.2 \leq$ [Fe/H] $\leq -0.5$. The atomic model for
Cr and details about NLTE calculations are documented in Sect.
\ref{sec:methods}. The statistical equilibrium of Cr under restriction of
different stellar parameters in discussed in Sect. \ref{sec:NLTE}. In Sect.
\ref{sec:Sun}, we present the NLTE analysis of solar \ion{Cr}{i} and
\ion{Cr}{ii} lines and derive the solar Cr abundance. In Sect. \ref{sec:mps}, we
describe observed spectra, stellar parameters, and present Cr abundances for a
sample of metal-poor stars. Finally, in Sect. \ref{sec:chem}, the revised
abundance ratios [Cr/Fe] are compared with the trends predicted by GCE models
and some implications for the evolution of Cr in the Galaxy are discussed.
\section{The methods}{\label{sec:methods}}
\subsection{Statistical equilibrium calculations}{\label{sec:atom}}
Restricted NLTE calculations for Cr are performed with the code DETAIL
\citep{Butler85}. In the more recent version of the code, radiative transfer is
based on the method of accelerated lambda iteration.

\begin{figure}
\centering
\resizebox{\columnwidth}{!}{\rotatebox{90}{\includegraphics{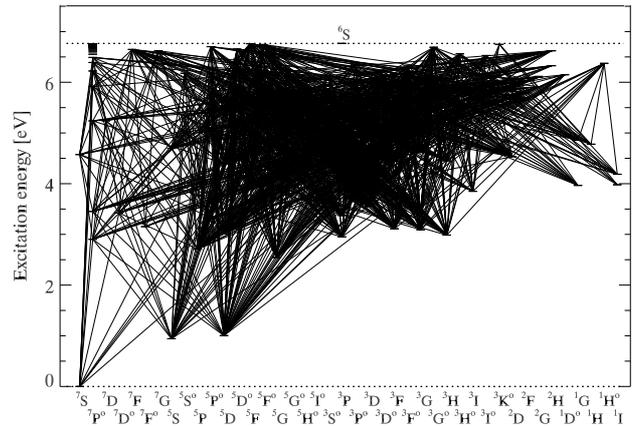}}}
\caption[]{Grotrian diagram of the \ion{Cr}{i} model atom. Lines represent
allowed transitions included in the model atom.}
\label{atom}
\end{figure}

The reference atomic model is constructed with $114$ levels for \ion{Cr}{i} and
$225$ levels for \ion{Cr}{ii}, with energies of $0.01$ eV and $1.1$ eV below the
respective ionization limits, $6.77$ eV and $16.49$ eV. The model is closed by
the \ion{Cr}{iii} ground state. The typical separation of fine structure
components in \ion{Cr}{i} and \ion{Cr}{ii} is less than $0.1$ eV. Hence, we do
not include fine structure in statistical equilibrium (SE) calculations. Each
term is represented by a single level with a total statistical weight of the
term and energy weighted by statistical weights of fine structure components.
Transitions are grouped. All transitions with oscillator strengths $\log gf \leq
-8$ and wavelengths $\lambda \geq 20000\ \AA$ are neglected. As a result, the
number of radiatively-allowed transitions is limited to $1590$ and $5216$ for
\ion{Cr}{i} and \ion{Cr}{ii}, respectively. Level excitation energies and
oscillator strengths were taken from the Kurucz online
database\footnote{http://kurucz.harvard.edu/}, supplemented by
the experimental data of Murray (1992). The Grotrian diagram of the \ion{Cr}{i}
model atom is shown in Fig. \ref{atom}. The atomic data for some levels,
including excitation energies E$_{\rm exc}$, wavelengths $\lambda_{\rm thr}$ and
cross-sections $\sigma_{\rm thr}$ at ionization thresholds, are given in Table
\ref{levels}.

\begin{figure}
\resizebox{\columnwidth}{!}{\rotatebox{90}
{\includegraphics{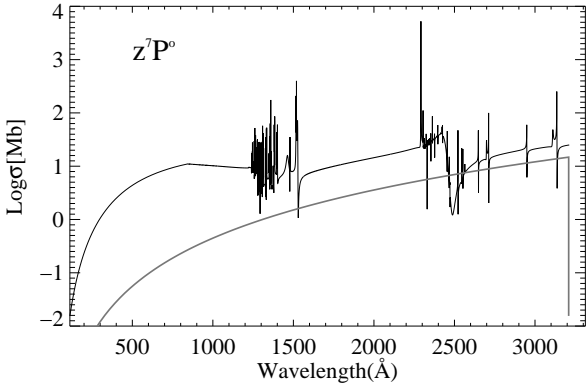}}}
\caption[]{Photoionization cross-section for the \Cr{z}{7}{P}{\circ}{} state 
of \ion{Cr}{i}: (black trace) quantum-mechanical calculations of Nahar (2009),
(gray trace) hydrogenic approximation with effective principle quantum number 
$n^* = 1.88$.}
\label{qm_photo}
\end{figure}

The following atomic processes are taken into account in calculations of
transition rates: radiative bound-bound and bound-free transitions, excitation
and ionization by collisions with free electrons, and neutral hydrogen atoms.
Scattering processes follow complete frequency redistribution. Recent
calculations of photoionization from quintet and septet states of \ion{Cr}{i}
\citep{2009JQSRT.110.2148N} revealed resonances in cross-sections, which are due
to the photoexcitation of the core leading to autoionization. The proper
treatment of the latter process, e.g. transitions to autoionizing states and
ionization into excited core states, is not yet possible with the DETAIL code.
Thus, we used partial cross-sections for ionization into the ground state of
\ion{Cr}{ii}, which are lower than the total cross-sections at energies above
the first excited state of the core, $\lambda \leq 1500~ \AA$. This is a
reasonable approximation for cool stellar atmospheres, because temperatures are
lower than those needed for core excitation. We have also shifted the
cross-sections to the observed edge energies of \ion{Cr}{i} states from the
NIST\footnote{http://physics.nist.gov/PhysRefData/} database, because the
latter are more accurate than the calculated LS term energies from
\citet{2009JQSRT.110.2148N}. Photoionization from the other \ion{Cr}{i} and
\ion{Cr}{ii} states is computed with a formula of Kramer
\citep{1935MNRAS..96...77M} corrected for the ion charge
\citep{2003ASPC..288...99R} and using the effective hydrogen-like principle
quantum number $n^*$. The use of effective quantum number increases the
cross-sections at the ionization edge for low-lying levels and reduces them for
some levels of high excitation that is more realistic than the original
hydrogenic approximation. The quantum-mechanical cross-sections $\sigma_{\rm
QM}(\nu)$ for the state \Cr{z}{7}{P}{\circ}{} are compared with the
hydrogenic approximation $\sigma_{\rm hyd}(\nu)$ in Fig. \ref{qm_photo}. The
cross-section at the photoionization edge, as well as the background, are larger
than $\sigma_{\rm hyd}(\nu)$. There are also strong resonances at energies,
where the solar atmospheric UV flux is sufficiently strong to produce the
overionization of the \Cr{z}{7}{P}{\circ}{} state. Quantum-mechanical
cross-sections for the other levels with excitation energies $E_{\rm exc}
\approx 3 - 4$ eV show a similar behaviour. We will show in Sect. \ref{sec:NLTE}
that these levels dominate ionization balance in \ion{Cr}{i}.

At present, there are no experimental and accurate theoretical data for
ionization and excitation in \ion{Cr}{i} by electron impact, as well as by
inelastic collisions with \ion{H}{i}. Thus, we rely on the commonly-used
approximations. Cross-sections for continuum and allowed discrete transitions
due to inelastic collisions with \ion{H}{i} were computed with the formulae of
\citet{1969ZPhy..225..470D}. This original recipe was developed for collisions
between equal hydrogen-like particles, hence we apply a scaling factor $0 \leq
\SH \leq 5$ to the Drawin cross-sections. The final choice of $\SH$ is discussed
in Sect. \ref{sec:abrat}. The rates of allowed and forbidden transitions due
collisions with electrons are calculated from the formulae of
\citet{1962ApJ...136..906V} and \citet{1973asqu.book.....A}, respectively. The
accuracy of the electron-impact cross-sections adopted in this work is not
better than that for collisions with neutral atoms \citep{1996ASPC..108..140M}.
However, the excitation and ionization balance in \ion{Cr}{i}/\ion{Cr}{ii}, as
well as abundances of Cr determined from the lines of both ionization stages,
depend only weakly on the exact treatment of inelastic $e^{-}$ collisions.
Our calculations show that even in the solar atmosphere ionization rates
due to collisions with \ion{H}{i} atoms dominate over the rates of ionization by
electrons for the majority of the Cr levels, exceeding the latter by nearly
three orders of magnitude for the uppermost levels. Moreover, collisional
ionization rates dominate over collisional excitation rates. In the atmospheres
of metal-poor stars, where the number density of free electrons is smaller than
in the Sun, collisional excitation in \ion{Cr}{i} is fully controlled by neutral
hydrogen.

In addition to the reference atomic model of Cr, which is described above, we
construct several test models. These models are used to show how the statistical
equilibrium of Cr changes under variation of different atomic parameters, like
size of the model atom, cross-sections for collisional and radiative
transitions.

\subsection{Model atmospheres and spectrum synthesis}

To maintain consistency in the abundance calculations, we have decided
to use the same type of model atmospheres, as employed for the derivation of
stellar parameters (see Sect. \ref{sec:mps}), microturbulence velocities, and
metallicites of the objects investigated in this work. These are classical
static 1D plane-parallel models MAFAGS-ODF without chromospheres
\citep{1997A&A...323..909F}. Line blanketing is treated with opacity
distribution functions from \citet{1992RMxAA..23...45K}. Convection is taken
into account with the mixing-length theory \citep{1958ZA.....46..108B} and the
mixing length is set to $0.5$ pressure scale heights. This value was chosen by
\citet{1993A&A...271..451F} to provide simultaneously the best fitting of Balmer
line profiles in the solar flux spectrum with $\Teff = 5780$ K.
\citet{2002A&A...385..951B} also suggest that the best fit of Balmer line
profiles can be achieved with $\alpha = 0.5$. Stratifications of temperature and
pressure in MAFAGS-ODF are similar to those given by other comparable models
\citep[see Fig. 15 in][]{2004A&A...420..289G}.

The abundances of Cr were computed by a method of spectrum synthesis with the
code SIU, kindly provided by T. Gehren (private communication). Standard line
broadening mechanisms, including radiation and quadratic Stark damping, were
taken into account. The line half-widths due to elastic collisions with
\ion{H}{i} are calculated using the cross-sections and velocity exponents
tabulated by \citet{1995MNRAS.276..859A}. In Table \ref{lines_cr}, they are
given in terms of van der Waals damping constants $\log C_6$, calculated for the
temperature $6000$ K.
\begin{table}
\small
\begin{minipage}{\columnwidth}
\tabcolsep1.3mm \caption{Selected levels of \ion{Cr}{i} in the model atom.
Photoionization cross-sections at the edge $\sigma_{\rm thr}$ are adopted from
Nahar (2009).} 
\label{levels}
\begin{tabular}{llrcl|llrcl}
\hline\noalign{\smallskip}
Level & g & E$_{\rm exc}$ & $\lambda_{\rm thr}$ & $\sigma_{\rm thr}$ & Level & g
& E$_{\rm exc}$ & $\lambda_{\rm thr}$ & $\sigma_{\rm thr}$ \\
      &   & eV            & $\AA$     & Mb\footnote{$1$ Mb $= 10^{-18}
{\rm cm}^2$} &       &   & eV            & $\AA$     & Mb \\
\noalign{\smallskip}\hline\noalign{\smallskip}
 \Cr{a}{7}{S}{}{}      & 7   & 0.00   &    1832  &   0.099 &   
\Cr{z}{5}{S}{\circ}{} & 5 & 5.35 &  8733  &   0.316  \\
 \Cr{a}{5}{S}{}{}      & 5   & 0.94   &    2128  &   0.316 &   
\Cr{e}{5}{D}{}{} & 25 & 5.46 &  9517  &   3.504  \\
 \Cr{a}{5}{D}{}{}      & 25 & 1.00   &    2151  &   0.327 &   
\Cr{w}{5}{P}{\circ}{} & 15 & 5.48 &  9645  &   4.793  \\
 \Cr{a}{5}{G}{}{}      & 45 & 2.54   &    2936  &   0.001 &   
\Cr{v}{5}{P}{\circ}{} & 15 & 5.57   &  10383  &   5.26  \\
 \Cr{a}{5}{P}{}{}      & 15 & 2.71   &    3055  &   2.133 &   
\Cr{f}{7}{S}{}{} & 7 & 5.66   &  11195  &   1.523  \\
 \Cr{z}{7}{P}{\circ}{} & 21 & 2.90   &    3209  &  25.02  &   
\Cr{f}{5}{S}{}{} & 5 & 5.7    &  11617  &   2.613  \\
 \Cr{z}{7}{F}{\circ}{} & 49 & 3.15   &    3431  &   0.911 &   
\Cr{f}{7}{D}{}{} & 35 & 5.8    &  12768  &   0.798  \\
 \Cr{z}{5}{P}{\circ}{} & 15 & 3.32   &    3599  &  27.3   &   
\Cr{u}{5}{P}{\circ}{} & 15 & 5.82   &  13091  &  29.07  \\
 \Cr{z}{7}{D}{\circ}{} & 35 & 3.42   &    3708  &   6.884 &   
\Cr{v}{5}{F}{\circ}{} & 35 & 5.91   &  14415  &  20.47  \\
 \Cr{y}{7}{P}{\circ}{} & 21 & 3.45   &    3741  &   0.89  &   
\Cr{g}{7}{D}{}{} & 35 & 5.92   &  14554  &  19.71  \\
 \Cr{y}{5}{P}{\circ}{} & 15 & 3.68   &    4014  &   3.756 &   
\Cr{w}{7}{P}{\circ}{} & 21 & 5.92   &  14565  &  38.33  \\
 \Cr{z}{5}{F}{\circ}{} & 35 & 3.85   &    4254  &   0.255 &   
\Cr{u}{5}{F}{}{} & 35 & 5.95   &  15143  &   0.028  \\
 \Cr{a}{5}{F}{}{}      & 35 & 3.89   &    4310  &   0.547 &   
\Cr{f}{5}{D}{}{} & 25 & 6.04   &  16944  &   4.298  \\
 \Cr{z}{5}{D}{\circ}{} & 25 & 4.17   &    4779  &   0.545 &   
\Cr{g}{7}{S}{}{} & 7 & 6.1    &  18526  &   2.968  \\
 \Cr{c}{5}{D}{}{}      & 25 & 4.40   &    5241  &  19.14  &   
\Cr{g}{5}{S}{}{} & 5 & 6.12   &   19033  &   4.949 \\
 \Cr{e}{7}{S}{}{}      & 7  & 4.57   &    5656  &   0.537 &   
\Cr{t}{5}{P}{\circ}{} & 10 & 6.17   &   20702  &  10.83   \\
 \Cr{e}{5}{S}{}{}      & 5  & 4.7    &    5990  &   0.903 &   
\Cr{s}{5}{F}{\circ}{} & 35 & 6.22   &   22667  &  40.11   \\
 \Cr{x}{5}{P}{\circ}{} & 15 & 5.08   &    7366  &  16.38  &   
\Cr{u}{5}{D}{\circ}{} & 25 & 6.28   &   25238  &   3.526  \\
 \Cr{y}{5}{F}{\circ}{} & 35 & 5.11   &    7468  &   0.091 &   
\Cr{h}{5}{S}{}{} & 5 & 6.33   &   28247  &   7.905  \\
 \Cr{y}{5}{D}{\circ}{} & 25 & 5.15 &  7687  &   1.267 &  
\Cr{t}{5}{D}{}{} & 25 & 6.45   &   39357  &   4.353  \\
 \Cr{z}{5}{H}{\circ}{} & 55 & 5.23 &  8088  &   0.009 &
\Cr{r}{5}{F}{}{} & 35 & 6.58   &   68020  &   0.129  \\ 
 \Cr{e}{7}{D}{}{}  & 35 & 5.24 &    8118  &   6.231 &
\Cr{h}{7}{D}{}{} & 32 & 6.62   &   86092  & 118.3  \\ 
 \Cr{x}{7}{P}{\circ}{} & 21 & 5.24 &  8119  &  27.37  &
\Cr{e}{7}{G}{}{} & 63 & 6.62   &   86731  & 151.0  \\
 \Cr{x}{5}{D}{\circ}{} & 25 & 5.29 &  8378  &  19.26  \\
\noalign{\smallskip}\hline\noalign{\smallskip}
\end{tabular}
\end{minipage}
\end{table}

\section{Statistical equilibrium of Cr}{\label{sec:NLTE}}

\begin{figure*}
\hbox{
\resizebox{8cm}{!}{{\includegraphics{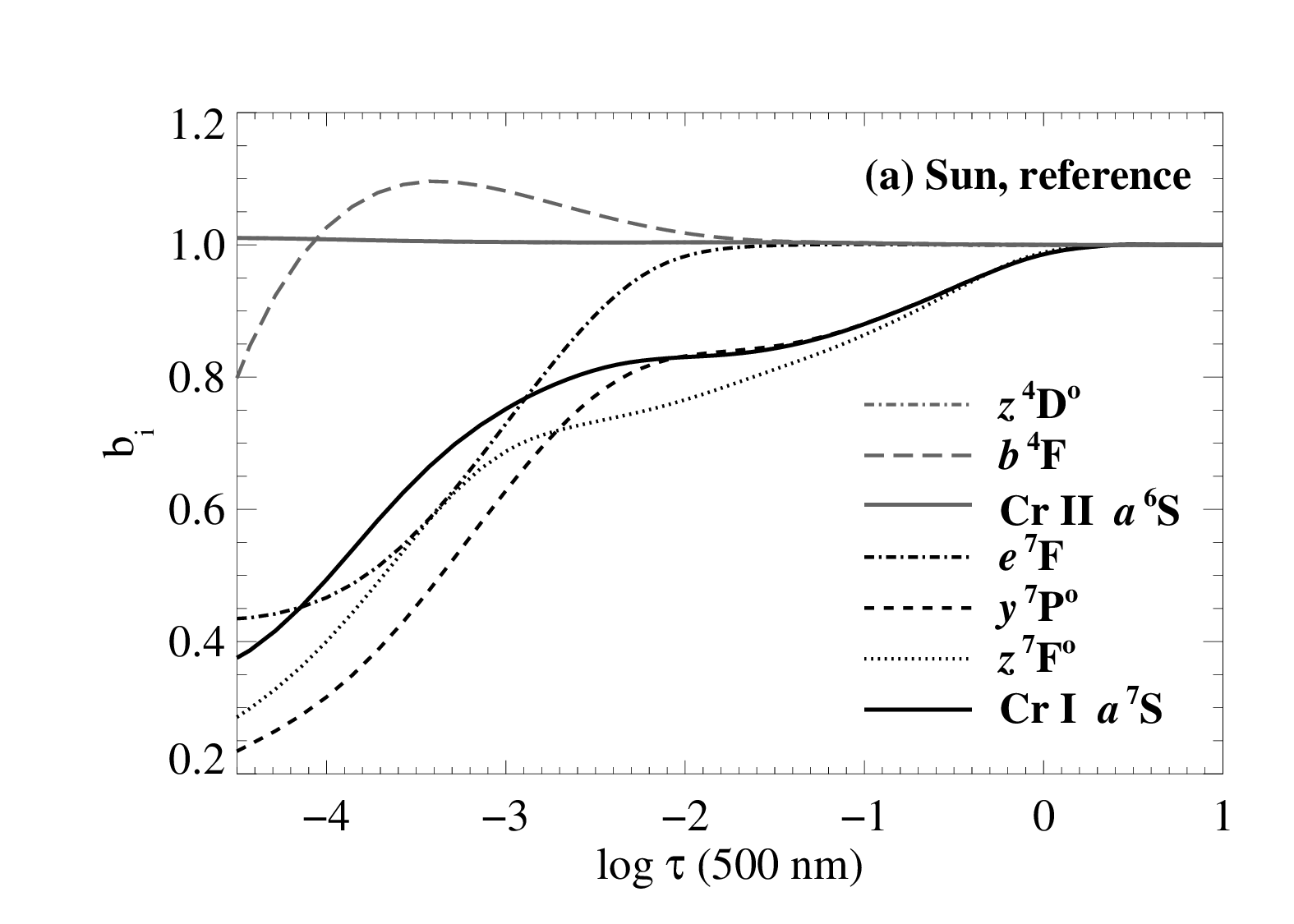}}}\hfill
\resizebox{8cm}{!}{{\includegraphics{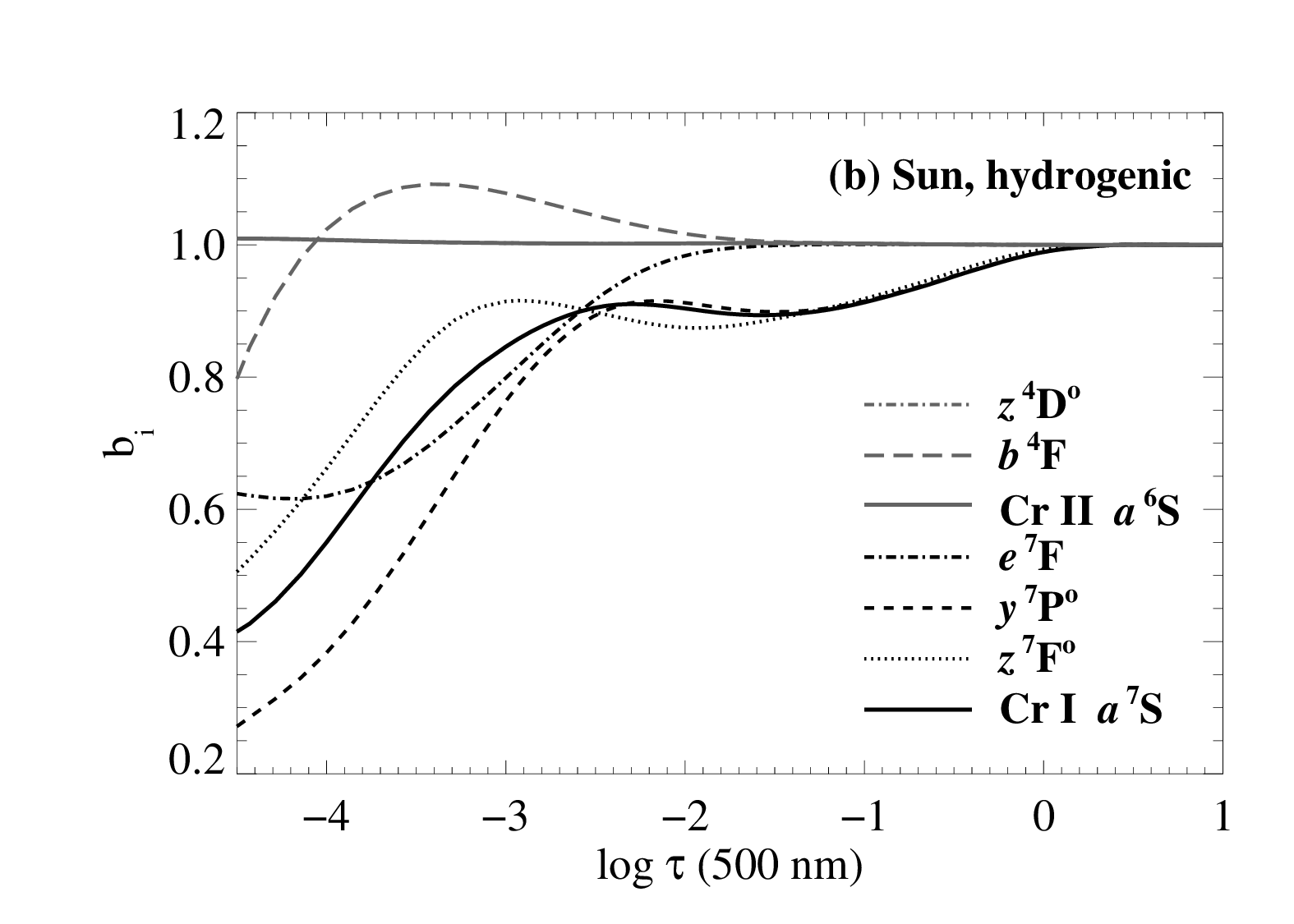}}}}
\vspace{0mm}
\hbox{\resizebox{8cm}{!}{{\includegraphics{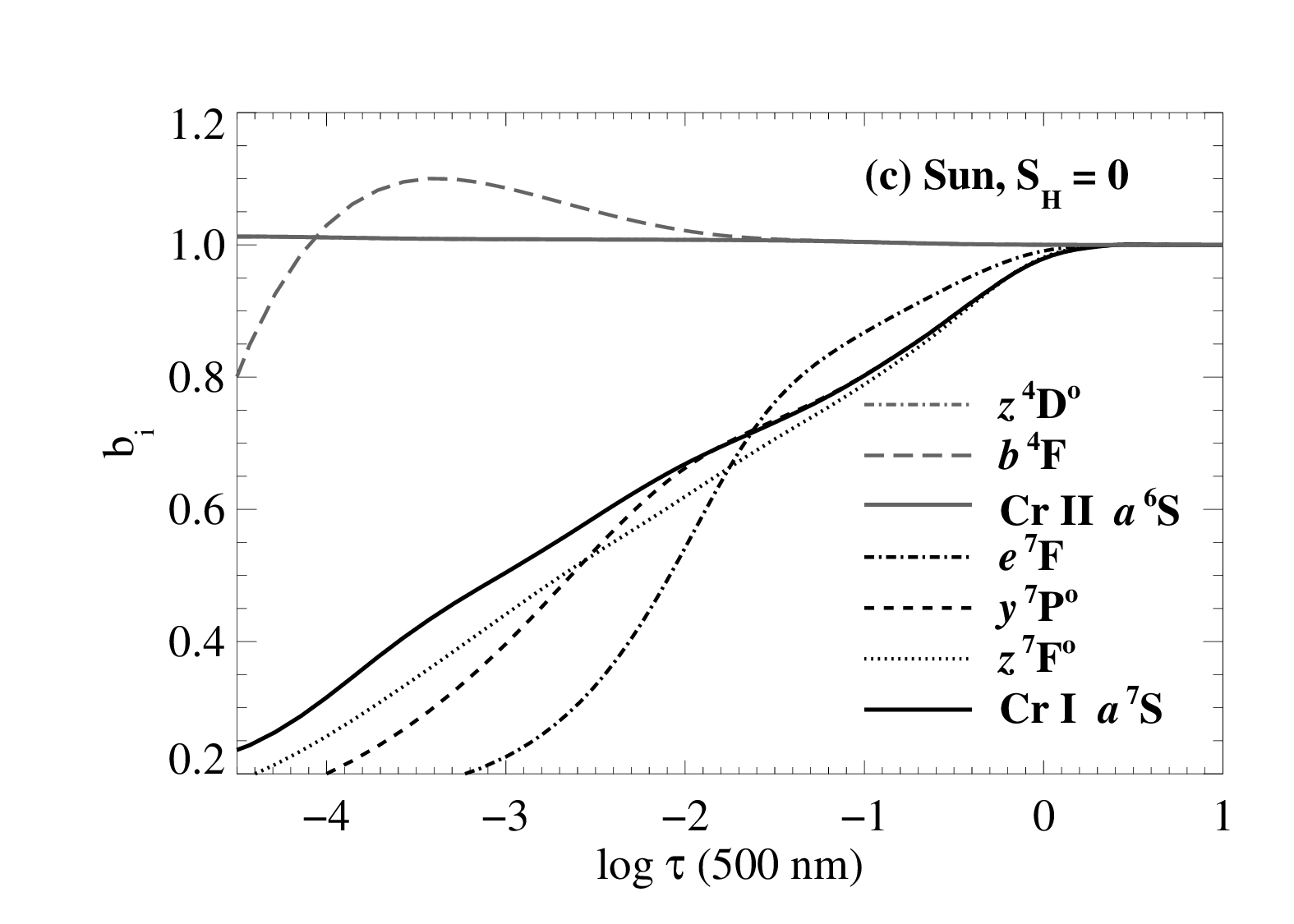}}}\hfill
\resizebox{8cm}{!}{{\includegraphics{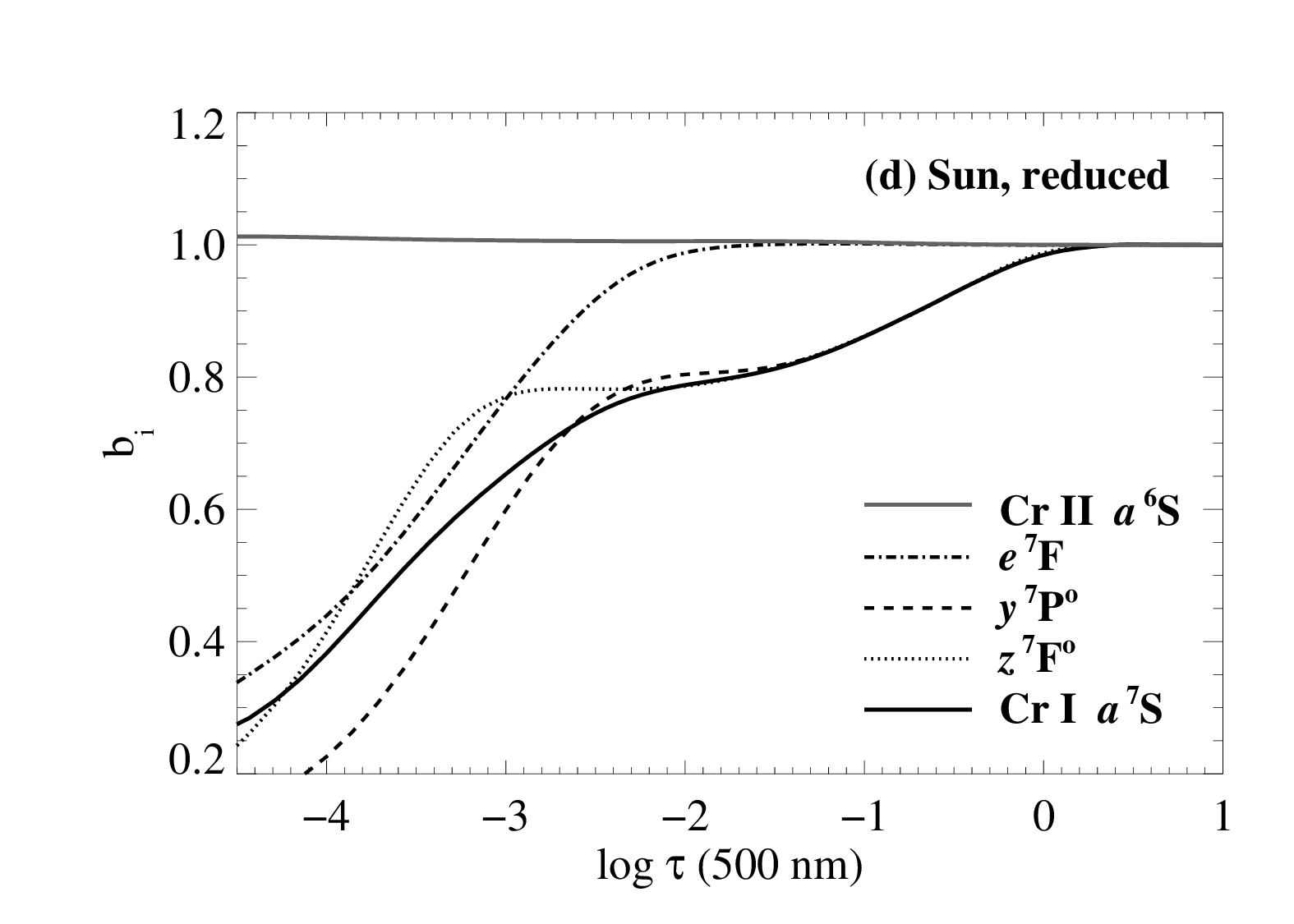}}}}
\vspace{0mm}
\hbox{\resizebox{8cm}{!}{{\includegraphics{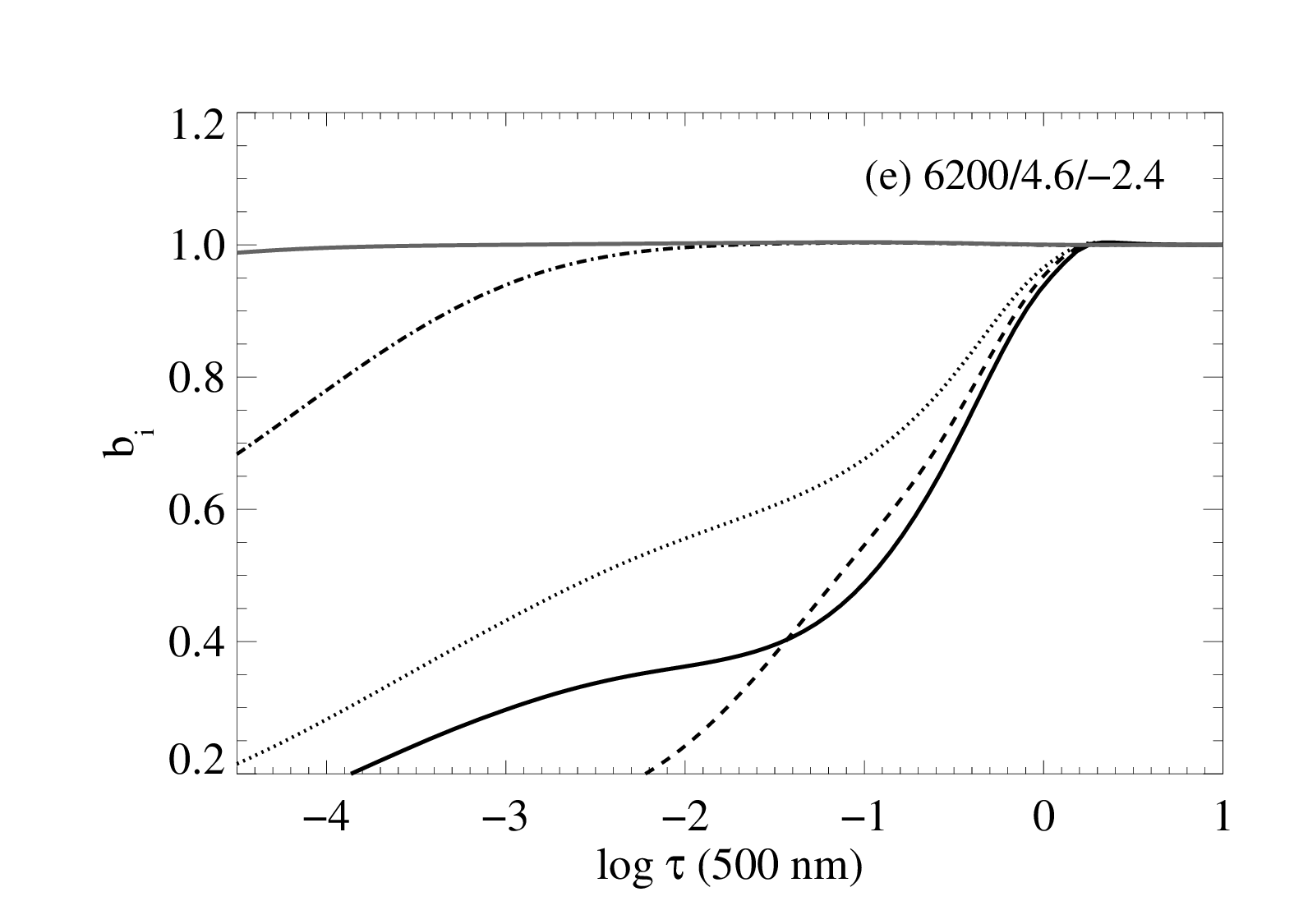}}}\hfill
\resizebox{8cm}{!}{{\includegraphics{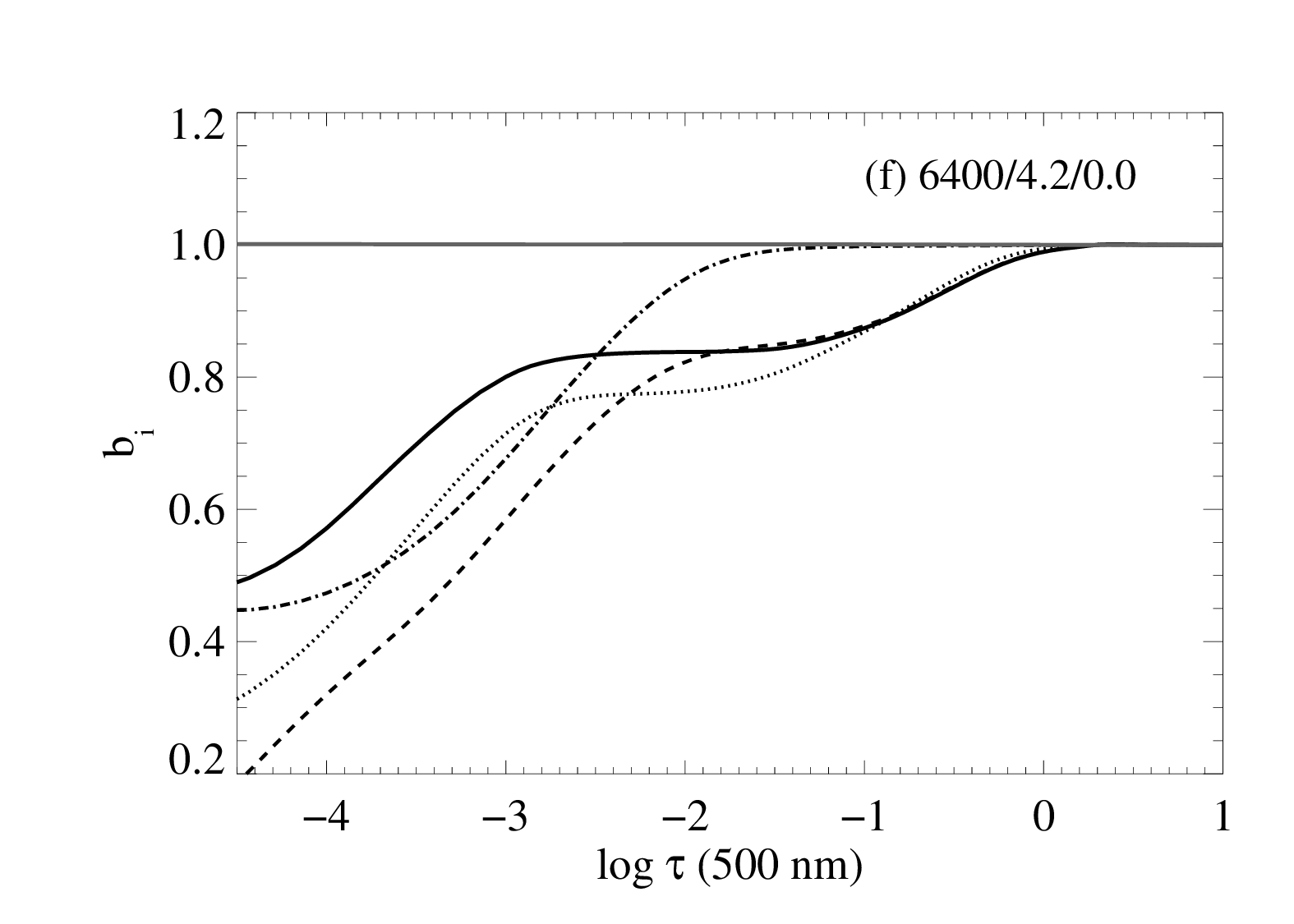}}}}
\vspace{0mm}
\hbox{\resizebox{8cm}{!}{{\includegraphics{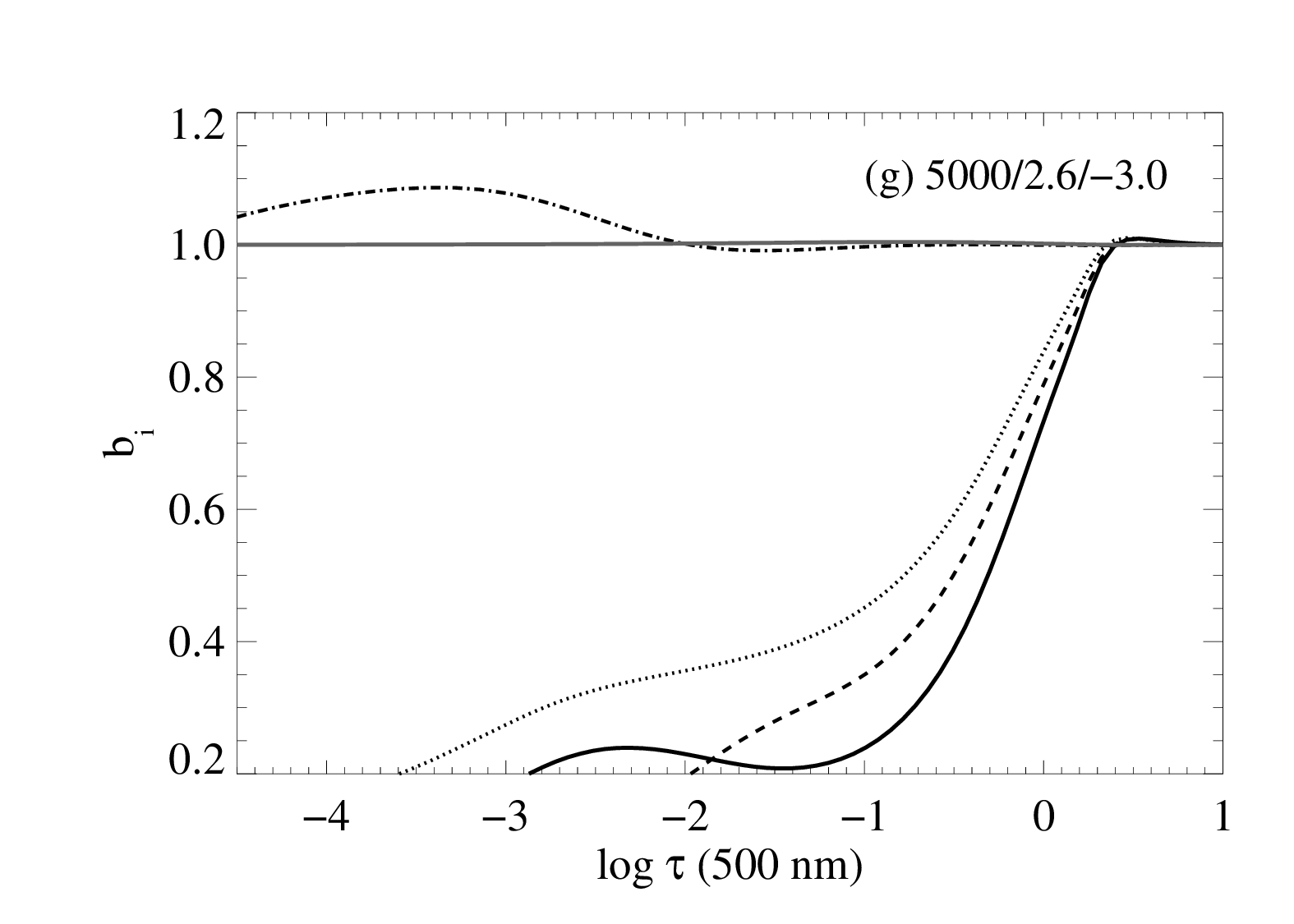}}}\hfill
\resizebox{8cm}{!}{{\includegraphics{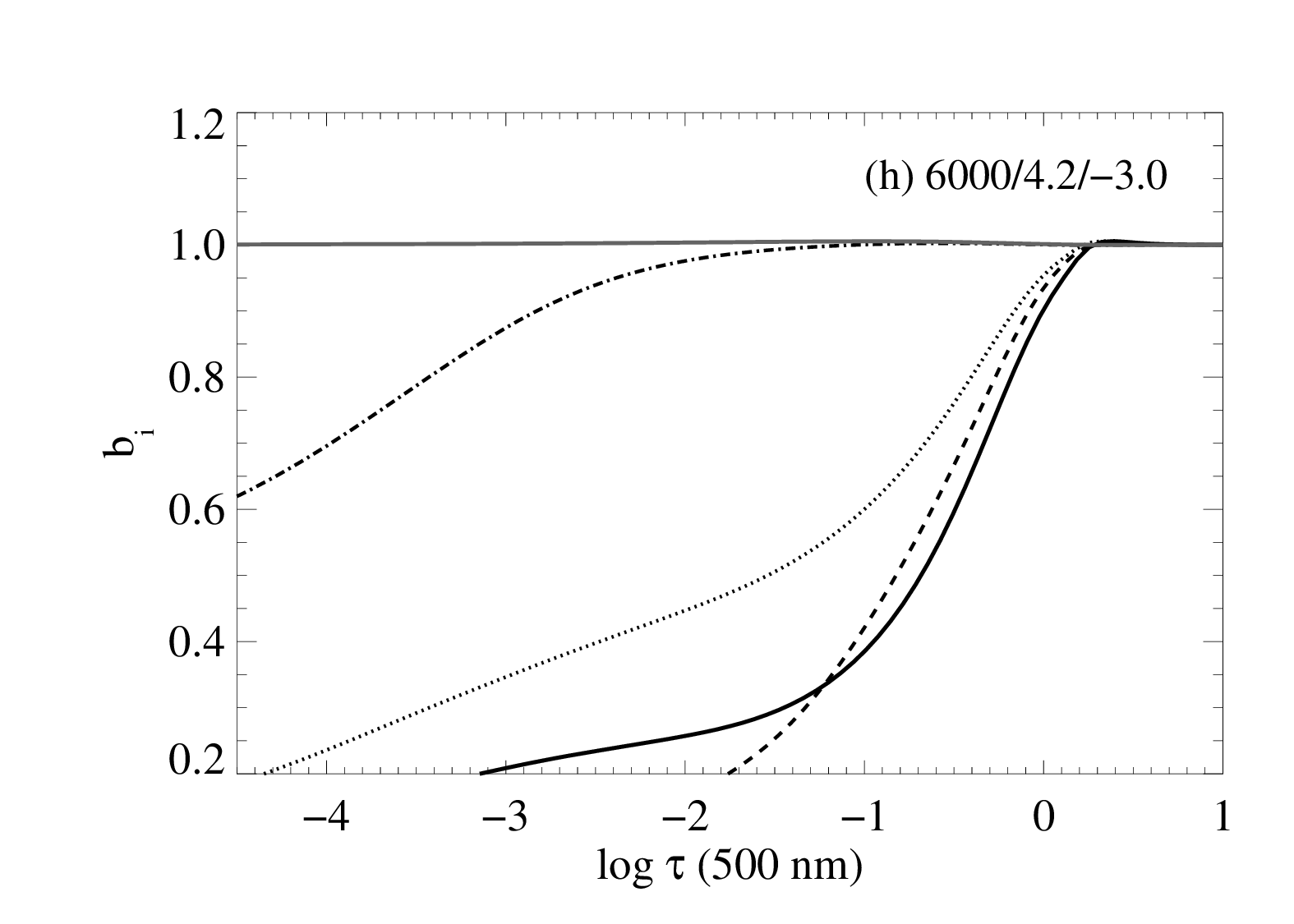}}}}
\caption[]{Departure coefficients $b_i$ of selected \ion{Cr}{i} and \ion{Cr}{ii}
levels as a function of optical depth at $500$ nm. Panels (a) to (d) present
results for the solar model atmosphere. (a): The reference model atom
with a total of $374$ levels constructed as described in Sect.
\ref{sec:atom}. Hydrogen collision rates are multiplied by $\SH = 0.05$.
Photoionization cross-sections for \ion{Cr}{i} are taken from Nahar (2009). (b):
Same as (a), but hydrogenic photoionization cross-sections are used for all
levels (c): Same as (a), but $\SH = 0$. (d): The reduced model atom with a total
of $108$ levels including the \ion{Cr}{ii} ground state, $\SH = 0.05$. (e) to
(h): Same as (a), but the model atmospheres are taken from the grid with stellar
parameters specified in each panel. Note that some levels on the plots
are in thermal equilibrium with the other levels.}
\label{cr_sun}
\end{figure*}

The departures of atomic level populations $n^{\rm{NLTE}}_i$ from their LTE
values $n^{\rm{LTE}}_i$ are best understood from the inspection of departure
coefficients, defined as $ b_i = n^{\rm{NLTE}}_i/n^{\rm{LTE}}_i $. The
departure coefficients for selected \ion{Cr}{i} and \ion{Cr}{ii} levels
calculated for the solar model atmosphere ($\Teff = 5780$ K, $\log g = 4.44$,
[Fe/H] = 0, $\Vmic = 0.9$ \kms) with different atomic models are shown as a
function of continuum optical depth $\opd$ at $500$ nm in Fig. \ref{cr_sun}.
We will mainly discuss NLTE effects in \ion{Cr}{i}, because the majority of
\ion{Cr}{ii} levels remain very close to LTE for the range of stellar parameters
we are interested in. As seen in the Grotrian diagram (Fig. \ref{atom}), Cr is a
very complex atomic system because a partially filled 3d electron shell 
results in a highly complex term structure with many low-lying and
closely-spaced energy levels. Hence, in contrast to simpler atoms like Li
\citep[Fig. 2]{2007A&A...465..587S} or Na \citep[Fig. 2]{1998A&A...338..637B},
clear isolation of transitions driving departures from LTE is not trivial. We
restrict the discussion of statistical equilibrium (SE) in Cr to few
simple cases, which will be illustrated with a few levels characteristic for
their depth dependency (Fig. \ref{cr_sun}).

The NLTE effects in \ion{Cr}{i} are similar to that of the other minority atoms
with complex configuration structure in cool stellar atmospheres, like
\ion{Fe}{i}, \ion{Mn}{i}, and \ion{Co}{i}. Deviations from LTE in \ion{Cr}{i}
develop in the layers where the mean intensity $J_{\rm \nu}$
exceeds the Planck function $B_{\nu}(\Te)$ over the bound-free edges of
significantly-populated low-lying \ion{Cr}{i} levels. Analysis of radiative
rates shows that the overionization is particularly strong from the odd levels
with large photoionization cross-sections, e.g. \Cr{z}{7}{P}{\circ}{}
($\lambda_{\rm thr} = 3209\ \AA$, Fig. \ref{qm_photo}), \Cr{z}{7}{F}{\circ}{}
($\lambda_{\rm thr} = 3431\ \AA$), and \Cr{z}{5}{P}{\circ}{} ($\lambda_{\rm thr}
= 3599\ \AA$). Thus, their $b_i$ drop below unity already at $\opd \sim +0.3$
(Fig. \ref{cr_sun}a). Overionization affects other low-excitation levels with
$E_{\rm exc} \sim 2 - 4$ eV, although the net ionization rates are by orders of
magnitude smaller then the rates from, e.g. \Cr{z}{7}{P}{\circ}{} level. The
cross-section of the ground state \Cr{a}{7}{S}{}{} shows a number of
autoionization resonances at $\lambda \leq 1700\ \AA$, where solar fluxes are
too low to produce significant ionization. The \ion{Cr}{i} ground state is
underpopulated via resonance transition in the multiplet $4$ (\Cr{a}{7}{S}{}{} -
\Cr{y}{7}{P}{\circ}{}), which is in detailed balance at $\opd \geq
-1.5$, and via collisions with the lowest metastable states, which are
overionized. A regular behaviour of departure coefficients for the majority of
\ion{Cr}{i} levels at $\opd \geq -2$ confirms the dominance of radiative
overionization at these depths with some deviations from relative thermal
equilibrium between low levels (compare levels \Cr{z}{7}{F}{\circ}{} and
\Cr{a}{7}{S}{}{} in Fig. \ref{cr_sun}a) reflecting the non-hydrogenic character
of their photoionization cross-sections. We have also performed SE
calculations with photoionization cross-sections for all \ion{Cr}{i} levels
derived from hydrogenic approximation. In this case, departure coefficients are
homogeneously distributed up to the depths $\opd \sim -1.5$ (Fig.
\ref{cr_sun}b); $b_i$ is a nearly monotonic function of the level ionization
potential.

Line transitions influence \ion{Cr}{i} excitation balance in the outer layers,
$\opd \leq -2$, where departure coefficients of the levels markedly deviate
from unity and from each other (Fig. \ref{cr_sun}a). Pumping by superthermal
radiation with $J_{\rm \nu} > B_{\nu}(\Te)$ becomes important in the layers,
where optical depth in the wings of strong low-excitation lines drops below
unity. Detailed balance in the transitions does not hold and their upper
levels are overpopulated. As soon as line cores form, $\tau_0^{\rm{l}} < 1$,
spontaneous transitions depopulate the upper levels and their $b_i$ steeply
decrease. The spectrum of \ion{Cr}{i} is represented by a large number of lines
in the near-UV, which are subject to these non-equilibrium excitation effects.
The transitions in the multiplet $4$, which sustain thermal equilibrium between
\Cr{a}{7}{S}{}{} and \Cr{y}{7}{P}{\circ}{} at $\opd \geq -1$, now go out of the
detailed balance and there is a small net radiative absorption leading to the
overpopulation of \Cr{y}{7}{P}{\circ}{} (Fig. \ref{cr_sun}a). At $\opd \sim
-3$, photon pumping ceases, because in the cores of the resonance lines
$\tau_0^{\rm{l}} < 1$, hence the departure coeficient of the
\Cr{y}{7}{P}{\circ}{} level drops. The levels with $E_{\rm exc} \sim 2.9 -
3.4$ eV are radiatively connected with the uppermost levels, which are separated
by an energy gap $\leq 0.5$ eV from the continuum, e.g. transitions 
\Cr{y}{7}{P}{\circ}{} - \Cr{e}{7}{F}{}{} with $\lambda \sim 3880$ \AA. Net
radiative rates in these transitions are positive at $\opd > -2$, but
overpopulation of the upper levels does not occur because they are strongly
coupled to the fully thermalized \ion{Cr}{ii} ground state. This coupling is
maintained by collisions with \ion{H}{i}, which exceed the respective rates of
ionization due to collisions with electrons by few orders of magnitude. There is
also a sequence of spontaneous de-excitations through the densely packed upper
levels with $E_{\rm exc} \sim 5 - 6$ eV, which is driven by $J_{\rm \nu} <
B_{\nu}(\Te)$ at the corresponding frequencies.

Fig. \ref{cr_sun}c demonstrates the behaviour of \ion{Cr}{i} level
populations, when inelastic collisions with \ion{H}{i} are neglected, $\SH = 0$.
Underpopulation of the levels at all optical depths is amplified, also the
uppermost levels, like \Cr{e}{7}{F}{}{}, decouple from the continuum already at
$\opd \approx 0$. Large differences with the reference model (Fig.
\ref{cr_sun}a), computed with $\SH = 0.05$, are seen at $\opd < -1.5$, where
electronic collisions are also not effective. Hence, we expect significant
changes in the opacity of the cores of strong \ion{Cr}{i} lines under NLTE. Also
at $\opd > -2$, departure coefficients of high-excitation levels are different
from the case of $\SH = 0.05$, thus affecting the source functions of
intermediate-strength lines.

Similarly, removal of the high-excitation \ion{Cr}{i} levels leads to slightly
increased deviations from LTE in \ion{Cr}{i}, although the variation of
departure coefficients with depth at $-2 < \opd < 0$ is not different from the
reference complete atomic model. Results for the reduced model atom, constructed
with $108$ levels of \ion{Cr}{i} with excitation energy of the highest level
$6.76$ eV and closed by the \ion{Cr}{ii} state, are shown in Fig. \ref{cr_sun}d.
This model is devoid of some doubly-excited \ion{Cr}{ii} states below the 1-st
ionization threshold and only transitions with $\log gf > -1$ are included. Note
the amplified underpopulation of the low and intermediate-excitation levels
compared to the reference model. This result can be easily understood as due to
the reduced radiative and collisional interaction of the \ion{Cr}{i} levels with
each other and with the continuum. Less electrons recombine and de-excite to the
lower levels via weak infrared lines, where $J_{\rm \nu} < B_{\nu}(\Te)$.

Statistical equilibrium of Cr in the atmospheres of cool subdwarfs and subgiants
is established by radiative processes. NLTE effects on the levels of \ion{Cr}{i}
and \ion{Cr}{ii} are amplified compared to the solar case. The main reason is
low abundances of metals, which supply free electrons and produce line
blanketing in the short-wave part of a spectrum. Hence, increased UV fluxes at
bound-free edges of low-excitation \ion{Cr}{i} levels lead to their strong
underpopulation, and collisional coupling between the levels is very weak due to
deficient electrons. As an example, we consider a metal-poor turnoff star with
[Fe/H] $= -2.4$ (Fig. \ref{cr_sun}e). Ionization balance is dominated by
radiative transitions from the low-excitation \ion{Cr}{i} levels. At $-1 < \opd
< 0.2$, these levels are depopulated by overionization and by optical pumping in
strong resonance lines, connecting the levels of \Cr{a}{7}{S}{}{} and
\Cr{y}{7}{P}{\circ}{} terms. At $\opd < -1$, photon pumping ceases because the
lines become optically thin, and spontaneous de-excitations maintain relatively
constant $b_i$ of the \Cr{a}{7}{S}{}{} state in the outer layers. An interesting
result is the presence of small deviations from LTE for intermediate-excitation
\ion{Cr}{ii} levels. As seen on the Fig. \ref{cr_sun}e, the odd level
\Cr{z}{4}{D}{\circ}{} with excitation energy $6.2$ eV is slightly overpopulated
by optical pumping at $-2 < \opd < 0$. But at smaller depths, photon losses in
transitions of the multiplet \Cr{b}{4}{F}{}{} - \Cr{z}{4}{D}{\circ}{} result in
a depopulation of the upper level. We use the lines of this multiplet in the
abundance analysis.

Effective temperature and gravity do not affect distribution of atomic level
populations at solar metallicity, but they become important with decreasing
[Fe/H]. The results for the model with $\Teff = 6400$ K, $\log g = 4.2$, and
[Fe/H]$= 0$ (Fig. \ref{cr_sun}f) are almost indistinguishable from those
obtained with the solar model atmosphere at optical depths $\opd >
-3$. This is stipulated by increased collisional interaction among the
\ion{Cr}{i} levels with $E_{\rm exc} > 3$ eV. Although the model flux maximum is
shifted to shorter wavelengths, ground state ionization is still not efficient
because the edge cross-section of the \Cr{a}{7}{S}{}{} state is very low. In
the cool model atmospheres, deviations from LTE depend on the stellar gravity,
and the effect is most pronounced at low metallicity. At [Fe/H] $= -3$,
departures from LTE for \ion{Cr}{i} levels are stronger in the model with $\Teff
= 5000$ K and $\log g = 2.6$ then in the model with $\Teff = 6000$ K and $\log g
= 4.2$ (Fig. \ref{cr_sun}g, h). This may account for a systematic difference
between metal-poor giants and dwarfs found by \citet{2008ApJ...681.1524L} and
\citet{2009A&A...501..519B} (see discussion in Sect. \ref{sec:abrat}).

The atomic model of Cr used in this work is not computationally
tractable in full NLTE calculations with 3D model atmospheres \citep[see
e.g.][for Ca]{BC99}, which require reduction of the model. Our analysis
suggests that it is possible to construct a simpler model of Cr atom, which
inherits the basic properties of the complete model and has similar performance
under restriction of different stellar parameters. In particular, removal of
high-excitation levels and numerous weak transitions in the \ion{Cr}{i} atom,
whose main effect is to provide stronger coupling of levels, can be compensated
by increasing collisional interaction between them. The carefully chosen scaling
factor to inelastic \ion{H}{i} collisions makes up for the missing transitions:
for \ion{Cr}{i}, the reduced model with $108$ levels and $\SH = 0.15$ gives a
similar description of the statistical equilibrium in Cr for solar-type stars to
the complete reference model with $340$ levels and $\SH = 0.05$.
However, there are two important concerns. First, the fact that the
simple model atom performs well in 1D does not guarantee that accurate results
are obtained with the same atomic model with 3D convective model atmospheres
\citep{2005ARA&A..43..481A}. Also, substituting the multitude of levels and
transitions in \ion{Cr}{i} by increased efficiency of collisions with \ion{H}{i}
(although both seem to produce the same effect on level populations) has
certainly no physical justification. In fact, our result even suggests that
large scaling factors to the Drawin's formula, as sometimes encountered in the
literature, may stem from the missing atomic data in SE calculations.

\section{Solar lines of \ion{Cr}{i} and \ion{Cr}{ii}}{\label{sec:Sun}}
All lines selected for the solar abundance analysis are given in Table
\ref{lines_cr}. We used the MAFAGS-ODF model atmosphere (Sect.
\ref{sec:methods}) with solar parameters. The profiles are broadened by the
solar rotation $V_{\rm rot , \odot} = 1.8$ \kms\ , microturbulence velocity
$\Vmic = 0.9$ \kms\ , and by a radial-tangential macroturbulence velocity
$\Vmac = 2.5 \ldots 4$ \kms. Radial and tangential components of $\Vmac$ are
assumed to be equal, and velocity distribution for each component is Gaussian.
$\Vmac$ is allowed to vary with the line strength and depth of formation. The
comparison spectrum was taken from the Solar Flux Atlas of
\citet{1984sfat.book.....K}.
\begin{table}
\begin{small}
\renewcommand{\footnoterule}{}  
\renewcommand{\tabcolsep}{2.5pt}
\caption{Lines of \ion{Cr}{i} and \ion{Cr}{ii} selected for solar and stellar
abundance calculations. Wavelengths $\lambda$ and excitation energies of
the lower levels of the transitions $\Elow$ are taken from NIST database. The
multiplet is specified in the 3-d column. Solar $\loggfe$ values for the
lines with an asterisk in the wavelength entry can not be reliably computed due
to severe blending. These lines are not used in determination of the solar Cr
abundance.}
\label{lines_cr}
\begin{center}
\begin{tabular}{llrrccrll}
\hline 
Id.$^a$ & $\lambda$ & Mult. & $\Elow$ & Lower & Upper & $\log gf$ & $\log C_6$ &
$\loggfe$ \\
 & ~~ [\AA] &  & [eV] & level & level & & & \\
\hline
\ion{Cr}{i} & & & & & & & & \\ 
bl & 4254.35* & 1 & 0.00 & \Cr{a}{7}{S}{}{3} & \Cr{z}{7}{P}{\circ}{4} &
-0.090 & -31.54 & - \\
bl & 4274.80* & 1 & 0.00 & \Cr{a}{7}{S}{}{3} & \Cr{z}{7}{P}{\circ}{3} &
-0.220 & -31.55 & - \\
bl & 4289.72* & 1 & 0.00 & \Cr{a}{7}{S}{}{3} & \Cr{z}{7}{P}{\circ}{2} &
-0.370 & -31.55 & - \\
bl & 4373.25 & 22 & 0.98 & \Cr{a}{5}{D}{}{2} & \Cr{z}{5}{F}{\circ}{1} & -2.300 &
-31.72 & 3.46 \\
 & 4492.31 & 197 & 3.38 & \Cr{b}{3}{P}{}{2} & \Cr{y}{3}{S}{\circ}{1} & -0.390 &
-31.34 & 5.34 \\
bl. w & 4496.86 & 10 & 0.94 & \Cr{a}{5}{S}{}{2} & \Cr{y}{5}{P}{\circ}{3} &
-1.140 & -31.62 & 4.61 \\
 & 4535.15 & 33 & 2.54 & \Cr{a}{5}{G}{}{3} & \Cr{z}{5}{G}{\circ}{4} & -1.020
& -31.59 & 4.74 \\
 & 4541.07 & 33 & 2.54 & \Cr{a}{5}{G}{}{4} & \Cr{z}{5}{G}{\circ}{3} & -1.150 &
-31.59 & 4.58 \\
 & 4545.96 & 10 & 0.94 & \Cr{a}{5}{S}{}{2} & \Cr{y}{5}{P}{\circ}{2} &
-1.370 & -31.63 & 4.39 \\
bl. w & 4591.39 &  21 & 0.97 & \Cr{a}{5}{D}{}{1} & \Cr{y}{5}{P}{\circ}{2}
& -1.740 & -31.75 & 4.01 \\
bl & 4600.75 & 21 & 1.00 & \Cr{a}{5}{D}{}{3} & \Cr{y}{5}{P}{\circ}{3} &
-1.250 & -31.74 & 4.49 \\
bl. w & 4613.37 & 21 & 0.96 & \Cr{a}{5}{D}{}{0} & \Cr{y}{5}{P}{\circ}{1} &
-1.650 & -31.76 & 4.13 \\
 & 4616.14 & 21 & 0.98 & \Cr{a}{5}{D}{}{2} & \Cr{y}{5}{P}{\circ}{2} &
-1.190 & -31.75 & 4.54 \\
 & 4626.19 & 21 & 0.97 & \Cr{a}{5}{D}{}{1} & \Cr{y}{5}{P}{\circ}{1} & -1.330 &
-31.76 & 4.42 \\
 & 4633.29 & 186 & 3.13 & \Cr{z}{7}{F}{\circ}{3} & \Cr{f}{7}{D}{}{4} & -1.110
& -31.18 & 4.62 \\
 & 4646.17 & 21 & 1.03 & \Cr{a}{5}{D}{}{4} & \Cr{y}{5}{P}{\circ}{3} & -0.740 &
-31.74 & 5.01 \\
bl. w & 4651.28 & 21 & 0.98 & \Cr{a}{5}{D}{}{2} & \Cr{y}{5}{P}{\circ}{1} &
-1.460 & -31.75 & 4.30 \\
bl. w & 4652.16 & 21 & 1.00 & \Cr{a}{5}{D}{}{3} & \Cr{y}{5}{P}{\circ}{2} &
-1.040 & -31.75 & 4.71 \\
 & 4700.61 & 62 & 2.71 & \Cr{a}{5}{P}{}{1} & \Cr{z}{5}{S}{\circ}{2} & -1.255 &
-31.61 & 4.51 \\
bl & 4708.04 & 186 & 3.17 & \Cr{z}{7}{F}{\circ}{5} & \Cr{f}{7}{D}{}{4} &
0.07 & -31.18 & 5.83 \\
bl. w & 4745.31 & 61 & 2.71 & \Cr{a}{5}{P}{}{3} & \Cr{x}{5}{D}{\circ}{4} &
-1.380 & -31.17 & 4.33 \\
 & 4801.03 & 168 & 3.12 & \Cr{a}{3}{F}{}{4} & \Cr{y}{3}{F}{\circ}{3} & -0.131 &
-31.35 & 5.62 \\
bl & 4885.78* & 30  & 2.54 & \Cr{a}{5}{G}{}{3} & \Cr{x}{5}{P}{\circ}{2} & -1.055 &
-31.3 & - \\
 & 4936.33 & 166 & 3.11  & \Cr{a}{3}{F}{}{3} & \Cr{z}{3}{H}{\circ}{4} & -0.25
& -31.37 & 5.47 \\
 & 4953.72 & 166 & 3.12 & \Cr{a}{3}{F}{}{4} & \Cr{z}{3}{H}{\circ}{4} & -1.48 &
-31.37 & 4.23 \\     
bl & 5204.52* & 7 & 0.94 & \Cr{a}{5}{S}{}{2} & \Cr{z}{5}{P}{\circ}{1} & -0.19
& -31.36 & - \\                                                                            
bl & 5206.04 & 7 & 0.94 & \Cr{a}{5}{S}{}{2} & \Cr{z}{5}{P}{\circ}{2} & 0.02
& -31.36 & 5.67 \\
bl & 5208.44 & 7 & 0.94 & \Cr{a}{5}{S}{}{2} & \Cr{z}{5}{P}{\circ}{3} & 0.17 &
-31.36 & 5.83 \\
 & 5241.46 & 59  & 2.71 & \Cr{a}{5}{P}{}{1} & \Cr{x}{5}{P}{\circ}{1} & -1.920
& -31.3 & 3.74 \\
bl. w & 5243.40 & 201 & 3.40 &  \Cr{z}{7}{D}{\circ}{2} &  \Cr{f}{7}{D}{}{1} &
-0.580 & -31.22 & 5.17 \\
 & 5247.56 & 18 & 0.96 & \Cr{a}{5}{D}{}{0} & \Cr{z}{5}{P}{\circ}{1} & -1.590 &
-31.41 & 4.28 \\
 & 5272.01 & 225 & 3.45 & \Cr{y}{7}{P}{\circ}{3} & \Cr{f}{7}{D}{}{4} & -0.420
& -31.2 & 5.31 \\
 & 5287.19 & 225 & 3.44 & \Cr{y}{7}{P}{\circ}{2} & \Cr{f}{7}{D}{}{3} & -0.870
& -31.21 & 4.84 \\
 & 5296.69 & 18 & 0.98 & \Cr{a}{5}{D}{}{2} & \Cr{z}{5}{P}{\circ}{1} &
-1.360 & -31.41 & 4.49 \\
 & 5300.75 & 18 & 0.98 & \Cr{a}{5}{D}{}{2} & \Cr{z}{5}{P}{\circ}{3} &
-2.000 & -31.41 & 3.73 \\
bl. w & 5304.21 & 225 & 3.46 & \Cr{y}{7}{P}{\circ}{4} & \Cr{f}{7}{D}{}{4} & 
-0.670 & -31.2 & 5.04 \\
 & 5312.88 &  225 &  3.45 &  \Cr{y}{7}{P}{\circ}{3} &  \Cr{f}{7}{D}{}{3} & 
-0.550 & -31.21 & 5.17 \\
 & 5318.78 &  225 &  3.44 &  \Cr{y}{7}{P}{\circ}{2} &  \Cr{f}{7}{D}{}{2} & 
-0.670 & -31.22 & 5.04\\
 & 5340.44 &  225 &  3.44 &  \Cr{y}{7}{P}{\circ}{2} &  \Cr{f}{7}{D}{}{1} & 
-0.730 & -31.22 & 5.02 \\
 & 5345.81 &  18  & 1.00 & \Cr{a}{5}{D}{}{3} & \Cr{z}{5}{P}{\circ}{2} &
-0.95 & -31.41 & 4.87 \\
 & 5348.32 &  18  & 1.00 & \Cr{a}{5}{D}{}{3} & \Cr{z}{5}{P}{\circ}{3} &
-1.21 & -31.41 & 4.60 \\
bl. w & 5409.79 &  18  & 1.03 & \Cr{a}{5}{D}{}{4} & \Cr{z}{5}{P}{\circ}{3} &
-0.670 & -31.41 & 5.15 \\
 & 5628.64 &  203 &  3.42 &  \Cr{b}{3}{G}{}{3} & \Cr{z}{3}{H}{\circ}{4} &
-0.740 & -31.34 & 4.99 \\
\ion{Cr}{ii} & & & & & & & & \\
 & 4555.00 &  44 & 4.07 & \Cr{b}{4}{F}{}{3} & \Cr{z}{4}{D}{\circ}{3} &
-1.25  & -32.4 & 4.42 \\
 & 4558.66 &  44 & 4.07 & \Cr{b}{4}{F}{}{4} & \Cr{z}{4}{D}{\circ}{3} &
-0.66  & -32.4 & 5.30 \\
bl & 4588.22 &  44 & 4.07 & \Cr{b}{4}{F}{}{3} & \Cr{z}{4}{D}{\circ}{2} &
-0.83  & -32.4 &  5.08 \\
 & 4592.09 &  44 & 4.07 & \Cr{b}{4}{F}{}{2} & \Cr{z}{4}{D}{\circ}{2} &
-1.42  & -32.4 &  4.47 \\
bl & 4634.11 &  44 & 4.07 & \Cr{b}{4}{F}{}{1} & \Cr{z}{4}{D}{\circ}{0} &
-0.98  & -32.4 &  4.69 \\
bl & 4812.35 &  30 & 3.86 & \Cr{a}{4}{F}{}{3} & \Cr{z}{4}{F}{\circ}{4} &
-2.10  & -32.46 & 3.73 \\
bl & 4824.14 &  30 & 3.87 & \Cr{a}{4}{F}{}{4} & \Cr{z}{4}{F}{\circ}{4} &
-0.92  & -32.46 & 5.01 \\
bl & 5237.34 &  43 & 4.07 & \Cr{b}{4}{F}{}{4} & \Cr{z}{4}{F}{\circ}{4} &
-1.16  & -32.4  & 4.51 \\
bl & 5308.44 &  43 & 4.07 & \Cr{b}{4}{F}{}{3} & \Cr{z}{4}{F}{\circ}{2} &
-1.81  & -32.4 &  3.84 \\
 & 5313.59 &  43 & 4.07 & \Cr{b}{4}{F}{}{2} & \Cr{z}{4}{F}{\circ}{2} &
-1.65  & -32.4 &  4.09 \\
\noalign{\smallskip}\hline\noalign{\smallskip}
\end{tabular}
\end{center}
$^a$ Line identification: ~bl - blended (in the line core or inner wing), ~bl. w
- blended in the wing only
\end{small}
\end{table}

Oscillator strengths for the \ion{Cr}{i} lines are taken from two sources. Most
of the data are from \citet{2007ApJ...667.1267S}, who determined branching
ratios of transitions by Fourier transform spectroscopy. The accuracy of $\log
gf$ values is within $5 - 10 \%$ for majority of transitions. In the absence of
data from Sobeck et al., we used transition probabilities from
\citet{1986MNRAS.220..303B} measured with the absorption technique.
\citet{FMW88} ascribe $10\%$ accuracy to these values. Oscillator strengths for
the majority of \ion{Cr}{ii} lines are taken from \citet{2006A&A...445.1165N}.
Branching ratios were measured with Fourier transform spectrometer and combined
with lifetimes from laser-induced fluorescence experiment. The uncertainty of
absolute oscillator strengths is $12 - 16\%$ for the transitions used in
this work, except for the \ion{Cr}{ii} line at $4592$ \AA\ with an uncertainty
of $37\%$. For the lines at $4555$ and $4812$ \AA, we used the $\log gf$'s from
\citet{1981JQSRT..25..167W}, and the \ion{Cr}{ii} lines of multiplet $43$ were
calculated with $\log gf$'s derived from the solar spectrum analysis by
\citet{1983AAfz...49...39K}. The accuracy of $\log gf$ values from
\citet{1981JQSRT..25..167W} and \citet{1983AAfz...49...39K} was later revised by
\citet{FMW88} to be of the order of $50\%$.

The LTE abundance of Cr determined from the \ion{Cr}{i} lines is $\loge = 5.66
\pm 0.04$ dex, where the uncertainty is one standard deviation. Our LTE
abundance is consistent with the meteoritic value, $5.63 \pm 0.01$
dex\footnote{The Cr abundance in CI-chondrites from \citet{2009arXiv0901.1149L}
was renormalized to the photospheric Si abundance of
\citet{2008A&A...486..303S}, $\log\varepsilon_{\rm Si, \odot} = 7.52$ dex}.
\citet{2007ApJ...667.1267S} calculated $\loge = 5.64 \pm 0.04$ dex from
equivalent widths of \ion{Cr}{i} lines in the disk-center intensity spectrum.
The agreement of our LTE abundance and that of Sobeck et al. is surprising,
since these authors used a semi-empirical Holweger-M\"{u}ller model atmosphere
and assumed a lower value of the microturbulence velocity, $\Vmic = 0.8$ \kms.
The abundances of Cr based on the LTE analysis of \ion{Cr}{ii} lines are similar
in our study and in Sobeck et al., $5.81 \pm 0.13$ dex and $5.77$ dex with
$\sigma = 0.13$ dex, respectively.

The LTE analysis of Cr lines revealed two problems. First, there is an
abundance discrepancy of $\sim 0.15$ dex between \ion{Cr}{i} and \ion{Cr}{ii}
lines, which also appears in the study of Sobeck et al. The line-to-line
abundance scatter for \ion{Cr}{ii} is also large, $\sigma = 0.13$ dex. Second,
there is a large spread of abundances within the multiplet $18$: profile fitting
of the \ion{Cr}{i} lines with equivalent widths $\EW > 60$ \mA\ requires
systematically higher abundances by $\sim 0.1$ dex compared to weaker lines. A
similar discrepancy was noted by \citet{1987A&A...180..229B}, who explained this
with non-thermal excitation in the multiplet $18$. Sobeck et al. did not support
the Blackwell's conclusion, although a $0.1$ dex line-to-line scatter in this
multiplet is also present in their results. We show in the next paragraph,
whether either of these two problems can be solved with NLTE.

The NLTE line formation is based on the departure coefficients $b_i$ computed
with the reference model of Cr atom (Sect. \ref{sec:methods}). We performed test
calculations for two scaling factors to inelastic collisions with hydrogen, $\SH
= 0$ and $0.05$. In both cases, NLTE corrections to abundances derived from
\ion{Cr}{i} lines are positive and range from $0.05$ to $0.1$ dex, depending on
the line strength and the excitation potential of the lower level. All
\ion{Cr}{i} lines computed under NLTE with $\SH = 0$ fit the observed spectrum
with the Cr abundance $\loge = 5.74 \pm 0.05$ dex, whereas a scaling factor $\SH
= 0.05$ leads to $\loge = 5.7 \pm 0.04$ dex. The NLTE abundance corrections to
the \ion{Cr}{ii} lines are small and negative. The results for $\SH = 0$ and
$\SH = 0.05$ are equal, $\loge = 5.79 \pm 0.12$ dex. In contrast to LTE, the
difference between both ionization stages is now fairly small, $\loge
(\rm{\ion{Cr}{ii}}) - \loge (\rm{\ion{Cr}{i}}) = 0.05$ dex for $\SH = 0$, and it
is within the combined errors of both values. At present, we can not
determine what causes the discrepancy between the abundance of Cr in meteorites
and our NLTE 1D abundance for the solar photosphere. The accuracy claimed for
oscillator strengths of \ion{Cr}{ii} and \ion{Cr}{i} transitions measured by
\citet{2006A&A...445.1165N} and \citet{2007ApJ...667.1267S}, respectively,
is very high. There is evidence that lines of a similar atom \ion{Fe}{i} are
affected by convective surface inhomogeneities, also NLTE effects on \ion{Fe}{i}
lines are different in 1D and in 3D radiative transfer calculations (see below).
For Cr, such detailed investigations are not yet available.

Our reference NLTE model satisfies ionization equilibrium of Cr, but it
does not solve the problem of overestimated abundances from several lines of
multiplet $18$. The solar NLTE abundance corrections\footnote{The difference in
abundances required to fit LTE and NLTE profiles is referred to as the NLTE
abundance correction, $\Delta_{\rm NLTE} = \log\varepsilon^{\rm NLTE} -
\log\varepsilon^{\rm LTE}$} $\Delta_{\rm NLTE}$ for all lines of this multiplet
are equal and they scale with the $\SH$ parameter. The behaviour of line
profiles under NLTE can be understood from the analysis of level departure
coefficients $b_i$ at the depths of line formation. The formation of weak lines
with $\EW < 60$ \mA\ is confined to $-1 < \opd < 0$, where $b_i = b_j$, so line 
source functions are thermal, $S_{\rm ij} = B_{\rm \nu}$. Hence, line intensity
profiles reflect the profile of an absorption coefficient, $\kappa_\nu \sim
b_i$. Since $b_i < 1$, the lines are weaker relative to their LTE strengths and
$\Delta_{\rm NLTE}$ is positive. Radiation in stronger lines of the multiplet
$18$ comes from the depths $-4 < \opd < 0$. The line cores have lower
intensities under NLTE due to depleted source functions, $S_{\rm ij} < B_{\rm
\nu}$, at $\opd < -2$. But their wings, which dominate the total line strength,
are formed at the depths where $b_i < 1$ due to overionization. The net effect
on the line profile is that the NLTE abundance correction is positive. For all
lines of multiplet $18$, we find $\Delta_{\rm NLTE} \approx +0.1$ dex and
$+0.05$ dex for $\SH = 0$ and $\SH = 0.05$, respectively. The result is
incompatible with the suggestion of \citet{1987A&A...180..229B}, since the
latter would require NLTE effects of different magnitude for all lines of
multiplet $18$. Moreover, in our 'extreme' NLTE model with $\SH = 0$, the net
NLTE effect on the profiles of stronger lines must be zero to eliminate the
abundance scatter. If this problem is due to a deficiency of the atomic model,
our main concern is the crude approximation used for inelastic collisions with
\ion{H}{i}. Compared to the Drawin's formulae, \textit{ab initio}
quantum-mechanical calculations \citep{1999PhRvA..60.2151B,2003PhRvA..68f2703B}
predict significantly lower collision rates for certain transitions of simple
alkali atoms, and they show that, in addition to excitation, other effects like
ion-pair formation become important. Thus, in principle, any rescaling of the
Drawin's cross-sections, e.g. using variable $\SH$ for levels of different
excitation energies as exemplified by \citet{1996A&A...307..961B}, does not
improve the reliability of results. Variation of $\SH$ does not reduce the
scatter among the \ion{Cr}{ii} lines, because the NLTE effects are very small
even for $\SH = 0$. For the same reason, employment of an LTE versus a
NLTE approach does not decrease the scatter.

\begin{figure}
\begin{center}
\resizebox{\columnwidth}{!}{\includegraphics[scale=1]{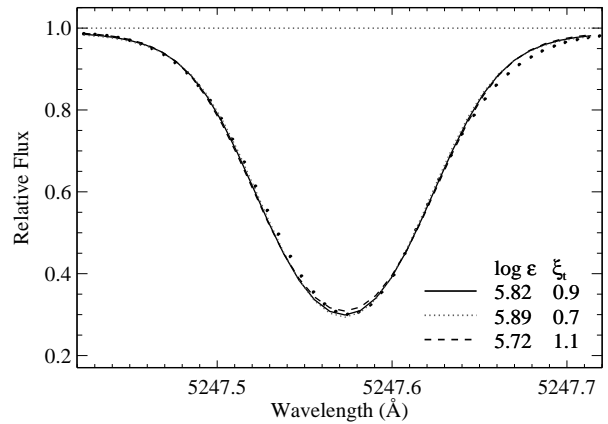}}
\caption{NLTE profiles of the \ion{Cr}{i} line at $5247$ \AA. Different 
combinations of microturbulent velocities $\Vmic$ and resulting Cr abundances
$\loge$ are shown. The line shows a convective asymmetry.} 
\label{line5247}
\end{center}
\end{figure}

The abundance anomaly among some \ion{Cr}{i} and \ion{Cr}{ii} lines may reflect
shortcomings of the 1D static mixing-length model atmospheres. Discrepant
\ion{Cr}{i} and \ion{Cr}{ii} lines with $70 < \EW < 120$ \mA\ are very
sensitive to a variation of the microturbulence parameter that was also
demonstrated for \ion{Fe}{i} lines with similar equivalent widths
\citep{2001A&A...380..645G}. For $\Delta \Vmic = \pm 0.2$ \kms\ with respect to
the reference value $0.9$ \kms, the abundances derived from the \ion{Cr}{i}
lines of multiplet $18$ change by roughly $\mp 0.1$ dex, and any combination of
$\Vmic$ and $\loge$ leads to an
equally good profile fit (Fig. \ref{line5247}). Using the depth-dependent
microturbulence as suggested by \citet{1967ZA.....65..365H} does not eliminate
the abundance scatter among \ion{Cr}{i} lines, although the \ion{Cr}{ii} lines
are strengthened requiring $0.1 - 0.15$ dex lower abundances.

Clearly, our approach of solving line formation in 1D homogeneous models with
a two-component gaussian velocity field is simple compared to multidimensional
radiative transfer calculations \citep[e.g.][]{2000A&A...359..669A}.
\citet{2002A&A...387..258S} have shown that abundance corrections due to
photospheric temperature fluctuations vary in sign and magnitude for lines of
different $\Elow$, $\lambda$, and $\EW$. For saturated lines, some effect due to
hydrodynamic velocity fields are also predicted. Besides, there is evidence that
NLTE effects on atomic level populations are amplified in the presence of
convective surface inhomogeneities \citep[e.g.][for the
Sun]{2001ApJ...550..970S}. Neutral iron is strongly overionized above hot
granules compared to cooler intergranular regions with the maximum effect on the
opacity of low-excitation \ion{Fe}{i} lines with $\Elow < 2$ eV. Deviations of
the source function from $B_{\nu}(\Te)$ are important for higher-excitation
lines in the intergranular regions. Shchukina \& Trujillo Bueno also find that
strong lines of any excitation potential formed above $\opd \sim -2.5$ are
sensitive to the variation of
collision rates. Comparable effects can be expected for the lines of
\ion{Cr}{i}, hence a systematic study of 3D effects in Cr is necessary to
confirm this. \citet{2009ARA&A..47..481A} computed LTE line formation for
\ion{Cr}{i} and \ion{Cr}{ii} with a hydrodynamical model atmosphere. Applying
NLTE abundance corrections derived from our 1D modelling, Asplund et al.
obtained the solar Cr abundance $5.64 \pm 0.04$ dex, which is almost equal to
our LTE 1D solar abundance determined from the \ion{Cr}{i} lines. Another study,
which provides an order-of-magnitude estimate of 3D effects on the Cr line
formation in very metal-poor atmospheres, is \citet{2009A&A...501..519B}. We
briefly discuss this paper in Sect. \ref{sec:abrat}.

Finally, we note that all problematic \ion{Cr}{ii} lines are located in the
spectral region $ 4500 - 4800$ \AA, where the continuum level is uncertain due
to severe blending. Hence, the abundance discrepancy may be, at least in part,
removed by adjusting the local continuum. On the other side, this procedure
could also affect \ion{Cr}{i} lines located in this spectral region.

\section{Abundances of Cr in metal-poor stars}{\label{sec:mps}}
\subsection{Observations and stellar parameters}
The list of stars along with their parameters is given in Table 
\ref{stel_param_gen}. Most of the stars were observed by T. Gehren and
collaborators with UVES spectrograph at the VLT UT2 on the Paranal, Chile, in
2001, and/or with the FOCES echelle spectrograph mounted at the 2.2m telescope
of the CAHA observatory on Calar Alto. The observations of HD 84937 were taken
from the UVESPOP survey \citep{2003Msngr.114...10B}. The UVES spectrum for G
64-12 was taken from ESO/ST-ECF Science Archive Facility (PID 67.D-0554(A)).
With the spectral resolution of $\lambda/\Delta\lambda \sim 40\,000 - 60\,000$,
it is not possible to distinguish the contribution of $\Vmac$ and $V_{\rm rot}$
to a line profile. Hence, the profiles are convolved with a Gaussian with the
full width at half maximum $\sim 2.8 - 4$ \kms.
\begin{table*}
\caption{Stellar parameters for the selected sample. NLTE and LTE
abundances of Cr computed using the lines of \ion{Cr}{i} and \ion{Cr}{ii} are
given as [Cr$_{\rm I}$/Fe] and [Cr$_{\rm II}$/Fe], respectively. The standard
deviations of these values are quoted as errors. The abundances of Fe are based
on \ion{Fe}{ii} lines. N$_{\rm Cr~I}$ and N$_{\rm Cr~II}$ are the number of
\ion{Cr}{i} and \ion{Cr}{ii} lines used for each star. See text for further
discussion of the data.} 
\renewcommand{\footnoterule}{} 
\label{stel_param_gen}
\begin{tabular}{lcccccrccccccc}
\hline\noalign{\smallskip}
Object & $\Teff$ & $\log g$ & $\xi_{\rm t}$ & [Fe/H] & N$_{\rm Cr~I}$ & N$_{\rm
Cr~II}$ & \multicolumn{2}{c}{[Cr$_{\rm I}$/Fe]} &
\multicolumn{2}{c}{[Cr$_{\rm II}$/Fe]} &
[Mg/Fe] \\
 ~ & K & & km/s & & & & NLTE & LTE & NLTE & LTE &  \\
\noalign{\smallskip}\hline\noalign{\smallskip}
HD 19445    & 5985 & 4.39 & 1.5 & $-1.96$ & 7 & 3 & $-0.07 \pm 0.05$ & $-0.3 \pm
0.06$ & $-0.09 \pm 0.01$ & $-0.11 \pm 0.03$ & 0.38 \\
HD 34328    & 5955 & 4.47 & 1.3 & $-1.66$ & 12 & 3 & $0.01 \pm 0.04$ & $-0.16
\pm 0.03$ & $0.02 \pm 0.02$ & $-0.01 \pm 0.02$ & 0.42 \\
HD 84937    & 6346 & 4.00 & 1.8 & $-2.16$ & 9 & 2 & $0.01 \pm 0.05$ & $-0.24 \pm
0.06$ & $-0.02 \pm 0.02$ & $-0.08 \pm 0.04$ & 0.32 \\
HD 102200   & 6120 & 4.17 & 1.4 & $-1.28$ & 16 & 4 & $0.01 \pm 0.06$ & $-0.11
\pm 0.06$ & $0.01 \pm 0.03$ & $0.0 \pm 0.04$ & 0.34 \\
BD$-4^\circ3208$ & 6310 & 3.98 & 1.5 & $-2.23$ & 6 & 0 & $0.04 \pm 0.03$ &
$-0.24 \pm 0.03$ & & & 0.34 \\
HD 122196   & 5957 & 3.84 & 1.7 & $-1.78$ & 10 & 2 & $-0.09 \pm 0.05$ & $-0.32
\pm 0.06$ & $-0.03 \pm 0.02$ & $-0.06 \pm 0.04$ & 0.24 \\
HD 123710   & 5790 & 4.41 & 1.4 & $-0.54$ & 14 & 4 & $-0.02 \pm 0.03$ & $-0.04
\pm 0.03$ & $-0.02 \pm 0.03$ & $-0.02 \pm 0.02$ & 0.24 \\
HD 148816   & 5880 & 4.07 & 1.2 & $-0.78$ & 21 & 6 & $0.01 \pm 0.05$ & $-0.06
\pm 0.07$ & $0.0 \pm 0.05$ & $0.0 \pm 0.04$ & 0.36 \\
HD 140283   & 5773 & 3.66 & 1.5 & $-2.38$ & 8 & 1 & $-0.03 \pm 0.06$ & $-0.38
\pm 0.04$ & $0.0$ & $-0.08$ & 0.43 \\
G 64-12     & 6407 & 4.20 & 2.3 & $-3.12$ & 4 & 0 & $0.13 \pm 0.04$ & $-0.26 \pm
0.02$ & & & 0.33 \\
\noalign{\smallskip}\hline\noalign{\smallskip}
\end{tabular}
\end{table*}

The sample includes nine objects from our recent studies
\citep{2008A&A...492..823B,2009MNRAS.tmp.1703B}, where the details on
observational material and derivation of stellar parameters can be found. 
In addition, we include here one thin disk star, HD 123710. Parameters of all
these stars are determined by \citet{2004A&A...413.1045G,2006A&A...451.1065G}
using the same MAFAGS-ODF model atmospheres, as employed in this work. The
effective temperatures were obtained from fitting the Balmer line profiles under
LTE. The spectroscopic temperatures are in agreement with the $\Teff$'s
determined by the method of Infrared Fluxes
\citep{1996A&AS..117..227A,1999A&AS..139..335A,2010arXiv1001.3142C}. The offset
with $\Teff$'s from the last reference is $\sim 20$ K with an \textit{rms} error
$23$ K. Surface gravities are based on \textit{Hipparcos} parallaxes.

Metallicities and microturbulence parameters were determined from LTE
fitting of \ion{Fe}{ii} lines by requiring a zero slope of Fe abundances versus
line equivalent widths. The validity of LTE approach in 1D line formation
calculations for \ion{Fe}{ii} was demonstrated in very accurate analyses of
statistical equilibrium of Fe in the Sun
\citep{2001A&A...366..981G,2001A&A...380..645G} and in metal-poor stars
\citep{2010IAUS..265..197M}. These studies employ the same type of atmospheric
models, as we use in the current work.

\citet{2004A&A...413.1045G,2006A&A...451.1065G} estimate the errors to be $100$
K for $\Teff$, $0.05$ dex for $\log g$, $0.05$ dex for [Fe/H], and $0.2$ km/s
for $\Vmic$. For the halo and thick disk stars, the model atmospheres were
computed with a total $\alpha$-element enhancement of $0.4$ dex. 

\subsection{[Cr/Fe] abundance ratios}{\label{sec:abrat}}
The abundances of Cr in metal-poor stars are calculated strictly relative to the
Sun. Any abundance estimate derived from a single line in a spectrum of
a metal-poor star is referred to that from the corresponding solar line, which
excludes the use of absolute oscillator strengths and of the average solar
abundance. We derive a differential element abundance for each detected Cr line
in a spectrum of a star according to $$ \mathrm{[El/H]}= \loggfestar -
\loggfesun $$ and then we average over all these lines for a given ionization
stage. Most of the Cr lines analyzed in the solar spectrum become very weak at
low metallicities, so typically we use from $3$ to $10$ lines for each
ionization stage.

\begin{figure}
\begin{center}
\resizebox{\columnwidth}{!}{\rotatebox{-90}{\includegraphics{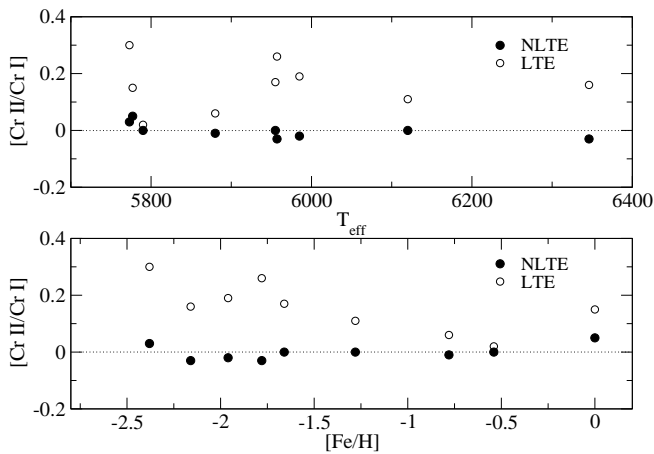}}}
\caption{[Cr/Fe] ratios as determined from \ion{Cr}{i} and \ion{Cr}{ii}
lines under NLTE (filled symbols) and LTE (open symbols) as a function of
effective temperature (top) and metallicity (bottom) for selected stars. The
NLTE abundances are computed with $\SH = 0$.} 
\label{cr_temp}
\end{center}
\end{figure}

Fig. \ref{cr_temp} displays the difference between [Cr/Fe] ratios computed from
the lines of two ionization stages, [Cr$_{\rm II}$/Cr$_{\rm I}$] $=$
[Cr$_{\rm II}$/Fe] - [Cr$_{\rm I}$/Fe], as a function of stellar effective
temperature and metallicity. NLTE and LTE ratios are shown with filled and open
symbols, respectively. NLTE abundances are based on SE calculations with the
atomic model of Cr with $\SH = 0$. There is an offset of $\sim 0.2 - 0.3$ dex
between \ion{Cr}{i} and \ion{Cr}{ii} lines under LTE, which increases with
decreasing [Fe/H] (Fig. \ref{cr_temp}, bottom). This reflects the fact that the
main stellar parameter that controls deviations from LTE in Cr is the metal
abundance. Since the NLTE effects of gravity and effective temperature are
smaller than that of [Fe/H], we do not see a clear trend with $\Teff$ in Fig.
\ref{cr_temp} (top), where no pre-selection according to $\log g$ or [Fe/H] was
made. A discrepancy of a similar magnitude and sign was reported by
\citet{2002ApJS..139..219J}, \citet{2008ApJ...681.1524L}, and
\citet{2009A&A...501..519B}. All studies agree that for giants with $\Teff \leq
5000$ the difference between two ionization stages is $\sim 0.4$ dex. For hotter
and higher-gravity models, characteristic of dwarfs, Lai et al. and Bonifacio et
al. derive smaller offsets, [Cr$_{\rm II}$/Cr$_{\rm I}$] $\sim 0.2$ dex, in good
agreement with our LTE calculations. These findings are consistent with
differential NLTE effects in Cr, which at low metallicity are larger for low
$\log g$ and low $\Teff$ models corresponding to giants (Sect. \ref{sec:NLTE}).

Our main result is that the offset between \ion{Cr}{i} and \ion{Cr}{ii} for any
combination of stellar parameters investigated here disappears when we use
atomic level populations computed with the reference atomic model with $\SH =
0$ (Fig. \ref{cr_temp}). This model allows for extreme NLTE effects on
\ion{Cr}{i} due to neglect of thermalizing \ion{H}{i} collisions. For $\SH =
0.05$, the abundances based on the \ion{Cr}{i} lines decrease by $\leq 0.05$
dex. Although these results are close to the abundances computed with $\SH =
0$, the agreement between two ionization stages is not so good, [Cr$_{\rm
II}$/Cr$_{\rm I}$] $\sim 0.04 \ldots 0.07$ dex, and the offset becomes even
larger with increasing $\SH$.

\begin{figure}
\begin{center}
\resizebox{\columnwidth}{!}{\rotatebox{-90}{\includegraphics{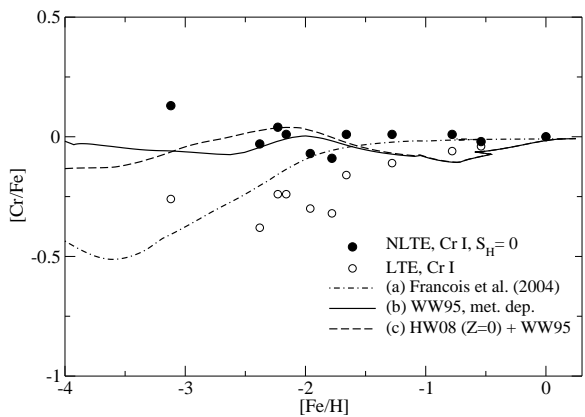}}}
\caption{Abundance ratios [Cr/Fe] as a function of metallicity. NLTE and
LTE-based Cr abundances in metal-poor stars are marked with filled and open
symbols. The evolutionary curves for [Cr/Fe] are computed with the CE model for
the solar neighborhood adopting different sets of SN II yields (a) - (c).} 
\label{cr_evol}
\end{center}
\end{figure}

The average Cr abundances for each star and their standard deviations are given in Table
\ref{stel_param_gen}. Table 4 with individual abundances for each line of \ion{Cr}{i} and \ion{Cr}{ii} is
available in the electronic edition of A\&A. [Cr$_{\rm I}$/Fe] and [Cr$_{\rm II}$/Fe] refer to the
analysis of \ion{Cr}{i} and \ion{Cr}{ii} lines, respectively. NLTE abundances
given in this table were computed with $\SH = 0$. In the spectra of
BD$-4^\circ3208$ and G 64-12 no lines of \ion{Cr}{ii} were detected, and
[Cr$_{\rm II}$/Fe] value for HD 140283 is based on one \ion{Cr}{ii} line.
In Fig. \ref{cr_evol}, [Cr/Fe] ratios computed under NLTE and LTE are plotted as
a function of the stellar iron abundance, with [Fe/H] based on the LTE analysis
of \ion{Fe}{ii} lines. The mean NLTE [Cr/Fe] ratio in stars with subsolar
metallicity is $\langle$[Cr/Fe]$\rangle$ $= 0$ with the dispersion $\sigma =
0.06$ dex. Assuming LTE, we derive $\langle$[Cr/Fe]$\rangle$ $= -0.21$ with
$\sigma = 0.11$ dex. The mean LTE value $\langle$[Cr/Fe]$\rangle$ and the slope
of [Cr/Fe] with [Fe/H] agree with other comparison studies, which refer to the
LTE analysis of \ion{Cr}{i} lines. \citet{2004A&A...416.1117C} showed that
[Cr/Fe] ratios in giants increase from $-0.5$ at [Fe/H] $ \approx -4$ to $-0.25$
at [Fe/H] $ \approx -2$, and the scatter in abundance ratios is very small,
$\sigma = 0.05$ dex. Such a small dispersion contradicts earlier results of
\citet{1995AJ....109.2757M} and \citet{1996ApJ...471..254R}, who demonstrate a
large spread of [Cr/Fe] abundances at any given metallicity. The reason for this
disagreement is that \citet{2004A&A...416.1117C} investigated only giant stars,
whereas the analysis of \citet{1995AJ....109.2757M} and
\citet{1996ApJ...471..254R} are based on a compilation of their own LTE
measurements in a mixed sample of stars and on other data from the literature.

The mean LTE abundances in metal-poor stars based on the \ion{Cr}{ii} lines are
identical in our study and in \citet{1991A&A...241..501G},
$\langle$[Cr/Fe]$\rangle = -0.05 \pm 0.04$ dex and $-0.04 \pm 0.05$ dex,
respectively. Both of these values are consistent with the mean NLTE abundance
derived from the \ion{Cr}{i} lines. This is not surprising, because we
find minor NLTE effects in \ion{Cr}{ii} for the range of stellar parameters
investigated here and excellent agreement is achieved between the \ion{Cr}{i}
and \ion{Cr}{ii} lines under NLTE.

Recently, \citet{2009A&A...501..519B} performed an LTE analysis of Cr abundances
in very metal-poor dwarfs and giants, demonstrating the influence of
multidimensional RT effects on \ion{Cr}{i} line formation in low-metallicity
model atmospheres. For giants, they determine a factor of $2$ smaller $3$D-$1$D
abundance corrections than for dwarfs. Although this result may explain the
systematic offset of $0.2$ dex, which they find between evolved and turnoff
stars, it is not clear how these 3D effects will manifest themselves in full 3D
NLTE RT calculations.

\section{Chemical evolution model for [Cr/Fe] ratio}{\label{sec:chem}}
\subsection{Nucleosynthesis}{\label{sec:nuc}}

Calculations of stellar nucleosynthesis and Galactic chemical evolution
indicate that the solar abundances of Fe and Cr derive for about $2/3$ from SNe
Ia and the residue from SNe II \citep{1986A&A...154..279M,1998ApJ...496..155S}.
In both instances, the formation of these even-$Z$ iron peak elements occurs in
explosive burning of silicon \citep{1999ApJ...511..862H,2007PrPNP..59...74T}.
$^{52}$Cr, which constitutes $83.8$\% of the solar Cr abundance, is formed as
radioactive $^{52}$Fe mainly in the incomplete Si-burning region
\citep{2002ApJ...565..385U,2005ApJ...619..427U}. The dominant isotope of Fe,
$^{56}$Fe ($91.7$\% of the solar Fe abundance) is produced as radioactive
$^{56}$Ni at very high temperatures, T $> 4\cdot10^{9}K$
\citep{1995ApJS..101..181W}. Since both $^{52}$Cr and $^{56}$Fe are formed as
$\alpha$-particle nuclei, the total yields of Cr and Fe are almost independent
of the neutron excess in the matter undergoing explosive burning and, to a first
order, of the metallicity of SN progenitors.

Although the nucleosynthesis sites for Cr and Fe are identified, theoretical
stellar yields of these elements are subject to large uncertainties
related to modelling the SN explosion. The amount of Fe-peak elements ejected in
a SN II primarily depends on the explosion energy and on the the mass
cut between the ejected material and the collapsed core (note that these
parameters are related, although they are often treated independently, e.g.
\citet{2002ApJ...565..385U}). Less Cr is produced relative to Fe for deeper
mass cut \citep{1999ApJ...517..193N,2002ApJ...565..385U}. The current crude
estimates of the mass cut are based on the mass ejected in the form of $^{56}$Ni
and on the ratio $^{58}$Ni/$^{56}$Ni, which are determined from SN II light
curves and spectra. Other parameters, which affect the relative
production of Fe-group nuclei, are the degree of mixing and fallback to the
remnant, energy and geometry of explosion
\citep{1999ApJ...517..193N,2002ApJ...565..385U,2005ApJ...619..427U}.
\citet{1997ApJ...486.1026N} demonstrated that an axisymmetric explosion in SN II
models enhances the production of $^{52}$Cr. All these parameters are
interlinked and depend on preceding stages of stellar evolution determined by
initial physical properties of a star.

The yields of Fe-peak elements from SNe Ia are also rather uncertain. It is not
clear what type of binary system evolution leads to an explosion, i.e. double or
single degenerate scenario, and whether the explosion proceeds via delayed
detonation or fast deflagration mechanism \citep[see][and references
therein]{1999ApJS..125..439I}. The production of neutron-rich isotopes in carbon
deflagration models depends on the neutron excess $\eta$, which in turn depend
on the ignition density and flame propagation speed. For exampple,
\citet{1999ApJS..125..439I} found that the yield of a neutron-rich $^{54}$Cr
isotope in the widely-used W7 model is very sensitive to $\eta$ in the
central region of a white dwarf model, where it is determined by electron
captures on nuclei. In this SN Ia model, the neutron-rich isotope of iron
$^{58}$Fe is produced in the outer regions where $\eta$ stems from CNO-cycle
production of $^{14}$N, i.e. the abundance of $^{58}$Fe is correlated with
initial metallicity of a white dwarf. Thus, depending on SN Ia model parameters,
different isotopic abundance ratios for Cr and Fe can be obtained. Some
fine-tuning of a SN Ia model may be necessary to reproduce the solar isotopic
composition.

\subsection{Model of chemical evolution}

The results we obtain for the chemical evolution of Cr for the solar
neighborhood are based on the model of \citet{2004A&A...421..613F}, which is a
revised version of the two-infall model presented by
\citet{1997ApJ...477..765C}. This model assumes two episodes for the Galaxy
formation: the first forms the halo on a short timescale of $\sim 1$ Gyr, the
second, much longer with a timescale of $\sim 7$ Gyr, produces the disc. The
initial mass function is adopted from \citet{1986FCPh...11....1S} and the star
formation law is the one proposed originally by \citet{1975ApJ...197..551T}.
Further details on the model can be found in \citet{2004A&A...421..613F}.

The following prescriptions for Fe-peak nucleosynthesis are adopted. For SN Ia,
we use the metallicity-independent yields of Cr and Fe computed by
\citet{1999ApJS..125..439I} with the updated version of Chandrasekhar mass W7
model of \citet{1984ApJ...286..644N}. This model predicts that $\sim 9$\% of the
total solar Cr is in the form of $^{54}$Cr. Hence, we have decreased the yield
of the neutron-rich isotope $^{54}$Cr by a factor of $4$ in order to reproduce
the solar abundance ratio of Cr isotopes. The overproduction of $^{54}$Cr can be
avoided in other SN Ia models with lower central ignition density and neutron
excess \citep{1999ApJS..125..439I}.
We use three sets of massive star yields. The first set (a) is
metallicity-independent yields from \citet{1995ApJS..101..181W} (WW95), computed
with the SN II models with a solar composition. We increased the Cr yields by a
factor of $10$ for the $10 - 20 M_{\odot}$ models following the recommendation
of \citet{2004A&A...421..613F}, who used this trick to avoid strong
underproduction of Cr relative to Fe with their CE model, [Cr/Fe]$ \approx -1$
at [Fe/H]$= -4$. We suggest, however, that the Cr/Fe deficit obtained by
\citet{2004A&A...421..613F} is not due to inadequate yields for low- and
intermediate-mass SNe II. Inspecting their Tables 1 and 2, we find very low Cr
and high Fe yields (from unknown source) for stars with $20 M_{\odot} \leq M
\leq 100 M_{\odot}$, which dominate Cr and Fe production at [Fe/H] $\leq -4$.
WW95 provide yields only for $M \leq 40 M_{\odot}$ models, and their Cr/Fe
production ratios for solar-metallicity stars with $M > 15 M_{\odot}$ are
supersolar.
In the second case (b), we adopt the metallicity-dependent WW95 yields for Cr
and Fe calculated for $Z/Z_{\odot} = 0$, $10^{-4}$, $10^{-2}$, $10^{-1}$, $1$,
and we perform linear interpolation between these values. The third set (c)
differs from the case (b) in that for $Z=0$ we use the yields of
\citet{2008arXiv0803.3161H}, computed for metal-free progenitors.

The [Cr/Fe] versus [Fe/H] relations computed with the chemical evolution models
(a) to (c) are compared with NLTE and LTE abundance ratios in Fig.
\ref{cr_evol}. The theoretical [Cr/Fe] ratios are normalized to the solar
abundances of Cr and Fe, predicted by each of the models. The Cr abundance in
the ISM at the solar system formation computed with the model (a) is $5.61$ dex,
whereas the models (b) and (c) predict $\log\varepsilon_{\rm Cr, \odot} = 5.68$
dex. The high abundance of Cr in the models (b) and (c) is compensated by a
higher Fe abundance $\log\varepsilon_{\rm Fe, \odot} = 7.53$.
The [Cr/Fe] trend based on LTE abundances derived from \ion{Cr}{i} lines is
qualitatively reproduced by the model (a) with adjusted WW95 yields. [Cr/Fe]
decreases with metallicity. This result is expected because our LTE abundances
generally agree with the spectroscopic abundances in metal-poor stars used by
\citet{2004A&A...421..613F} to calibrate SN yields in their model. The flat
[Cr/Fe] trend based on NLTE abundances of Cr is well followed by the model (b).
Deviations of [Cr/Fe] from zero are due to mass and metallicity-dependence of
yields in the WW95 models. The Cr/Fe production ratios are supersolar for more
massive stars, $M \geq 18 M_{\odot}$, with $Z/Z_{\odot} = 0.01, 0.1, 1$.
However, the IMF-integrated Cr/Fe production ratios are subsolar for SNe II with
$Z = 0.01$ and $Z = 0.1$ that is reflected in declining [Cr/Fe] for
metallicities $-2 <$ [Fe/H] $< -1$. At [Fe/H] $> -1$, [Cr/Fe] rises again due to
supersolar Cr/Fe production from exploded massive stars with initial solar
metallicity, $Z = Z_{\odot}$. The total solar Cr abundance is attained due to
dominant contribution ($\sim 60$\%) of SNe Ia to production of this element. 
We obtain similar results using the new nucleosynthesis yields of
\citet{2008arXiv0803.3161H} for metal-free stars, model (c) in Fig.
\ref{cr_evol}. The difference with model (b) can be seen only at very low
metallicity, [Fe/H] $< -3$.

Evolutionary curves for [Cr/Fe], which are similar to the models (b) and (c),
were calculated by \citet{1995ApJS...98..617T,1998ApJ...496..155S,
2000A&A...356..873A,2000A&A...359..191G, 2006ApJ...653.1145K}. However, good
agreement between our NLTE [Cr/Fe] trend and predictions of different
theoretical GCE models does not allow us to constrain the latter. The reason is
that different studies adopt different prescriptions for stellar
nucleosynthesis, mixing in the ISM, recipes for inflow and outflow, IMF, etc. We
are not aware of any complete systematic study in the literature, focusing on
the effect of these and other parameters on evolution of elemental abundance
ratios. Studies analogous to that of \citet{2005A&A...430..491R}, who
investigated the effect of initial mass function and stellar lifetimes, are
highly desirable.

\section{Conclusions}

The atomic level populations of Cr in the atmospheres of late-type stars are
very sensitive to the non-local UV radiation field, which drives them away from
the LTE distribution given by Saha-Boltzmann statistics. Using 1D static
model atmospheres, we find strong NLTE effects for the minority ion \ion{Cr}{i},
they are related to overionization from the low-excitation odd \ion{Cr}{i}
levels with large quantum-mechanical photoionization cross-sections. The NLTE
abundance corrections to \ion{Cr}{i} lines are positive and increase with
decreasing model metallicity and gravity. The number densities of excited states
in \ion{Cr}{ii} are also modified by non-equilibrium excitation processes, but
the effect on abundances is negligible for dwarfs and subgiants analyzed in
this work.

Ionization equilibrium of Cr for the Sun and the metal-poor stars is satisfied,
if we allow for extreme NLTE effects in \ion{Cr}{i} by neglecting inelastic
collisions with \ion{H}{i} in statistical equilibrium calculations. For the Sun,
the NLTE Cr abundance is $5.74 \pm 0.04$ dex, it is $0.05$ dex lower than that
computed from \ion{Cr}{ii} lines. Both values disagree with the Cr abundance in
\ion{C}{i} meteorites, $5.63 \pm 0.01$ dex (this value was taken from
\citet{2009arXiv0901.1149L} and was renormalized to the photospheric Si
abundance of \citet{2008A&A...486..303S}). A few \ion{Cr}{i} and \ion{Cr}{ii}
lines give systematically higher abundances. Since these lines are sensitive to
microturbulence parameter, the abundance anomaly most likely reflects the
shortcomings of our 1D model atmospheres. Line formation calculations with 3D
hydrodynamical models are necessary to confirm this.

The NLTE \ion{Cr}{i}-based abundances in metal-poor stars are
systematically larger than that computed under LTE approach. The difference of
$0.2 - 0.4$ dex is due to substantial overionization of \ion{Cr}{i} at low
metallicity. The LTE abundances determined in this work using \ion{Cr}{i} lines
are consistent with other LTE studies \citep{2002ApJS..139..219J, 2004A&A...416.1117C, 2004ApJ...612.1107C, 2008ApJ...681.1524L,
2009A&A...501..519B}, confirming that declining [Cr/Fe] with metallicity is an
artifact of LTE assumption in line formation calculations. The mean NLTE [Cr/Fe]
ratio in stars with subsolar metallicity computed from \ion{Cr}{i} lines is
$\langle$[Cr/Fe]$\rangle$ $= 0$ with the standard deviation $\sigma = 0.06$ dex.
Using the \ion{Cr}{ii} lines, we derive $\langle$[Cr/Fe]$\rangle = -0.05 \pm
0.04$ dex. The finding that [Cr/Fe] remains constant down to lowest
metallicities is consistent with nucleosynthesis theory, which predicts that Cr
and Fe are co-produced in explosive Si-burning in supernovae.

The NLTE [Cr/Fe] trend with [Fe/H] is reproduced by most of the Galactic
chemical evolution models, without the need to invoke peculiar conditions in the
ISM or to adjust theoretical stellar yields.  Our theoretical evolution of
[Cr/Fe] in the solar neighborhood, computed with the two-infall GCE model of
\citet{1997ApJ...477..765C} with SN Ia yields from \citet{1999ApJS..125..439I}
and metallicity-dependent SN II yields from \citet{1995ApJS..101..181W}, is in
agreement with the NLTE results. The model predicts that $\sim 60$\% of
the total solar Cr and Fe are due to SNe Ia and the rest due to SNe II, the
latter synthesize both elements in roughly solar proportions. The
underproduction of Cr relative to Fe in SNe Ia is compensated by its
overproduction in solar-metallicity SNe II.

\begin{acknowledgements}
Based on observations made with the European Southern Observatory telescopes
(obtained from the ESO/ST-ECF Science Archive Facility) and the Calar Alto
Observatory telescopes. MB thanks Dr. Aldo Serenelli for useful comments on
nucleosynthesis of Cr and critical revision of the manuscript.
\end{acknowledgements}
\bibliographystyle{aa}
\bibliography{references}
\end{document}